\documentclass[preprint,showpacs,showkeys,preprintnumbers,prb]{revtex4}

\usepackage{graphicx,epsfig}

\begin{document}

\title{Correlation in the transition metal based Heusler compounds Co$_2$MnSi and Co$_2$FeSi}

\author{Hem Chandra Kandpal, Gerhard H. Fecher, and 
Claudia Felser} 
\email{felser@uni-mainz.de}
\affiliation{
Institut f\"ur Anorganische und Analytische Chemie,
Johannes Gutenberg - Universit\"at, D-55099 Mainz, Germany.}
\author{Gerd Sch\"onhense}
\affiliation{
Institut f\"ur Physik,
Johannes Gutenberg - Universit\"at, D-55099 Mainz, Germany.}

\date{\today}

\begin{abstract}

Half-metallic ferromagnets like the full Heusler compounds with formula X$_2$YZ are supposed to show an integer value of the spin magnetic moment. Calculations reveal in certain cases of X = Co based compounds non-integer values, in contrast to experiments. In order to explain deviations of the magnetic moment calculated for such compounds, the dependency of the electronic structure on the lattice parameter was studied theoretically. In local density approximation (LDA), the minimum total energy of Co$_2$FeSi is found for the experimental lattice parameter, but the calculated magnetic moment is about 12\% too low. Half-metallic ferromagnetism and a magnetic moment equal to the experimental value of $6\mu_B$ are found, however, only after increasing the lattice parameter by more than 6\%.

To overcome this discrepancy, the LDA$+U$ scheme was used to respect on-site electron correlation in the calculations. Those calculations revealed for Co$_2$FeSi that an effective Coulomb-exchange interaction $U_{eff}=U-J$ in the range of about 2eV to 5eV leads to half-metallic ferromagnetism and the measured, integer magnetic moment at the measured lattice parameter. Finally, it is shown in the case of Co$_2$MnSi that correlation may also serve to destroy the half-metallic behavior if it becomes too strong (for Co$_2$MnSi above 2eV and for Co$_2$FeSi above 5eV). These findings indicate that on-site correlation may play an important role in the description of Heusler compounds with localized moments.

\end{abstract}



\maketitle

\section{Introduction}

Half-metallic ferromagnetism was first predicted for the half Heusler compound NiMnSb by de Groot {\it et al} \cite{GME83}. A half-metallic ferromagnet is a material with a gap between spin states at the Fermi energy ($\epsilon_F$), such that the majority states are conducting and the minority states are semi-conducting. Half-metallic ferromagnets (HMF) are supposed to exhibit a real gap in the minority (or majority) density of states (DOS). This gap has in compounds with integer site occupancies the consequence that the magnetic moment has to be an integer, because the number of occupied minority states is integer. It should be noted that the HMF character is lost if an already small deviation from an integer value is observed.

Certain X$_2$YZ Heusler compounds have also been predicted to show half-metallic ferromagnetism \cite{IFK95}. Co and Mn containing Heusler compounds have attracted particular attention, as they are strongly ferromagnetic with high Curie temperatures (above 900K) \cite{BNW00}. One of them - Co$_2$MnSi - is an attractive candidate for many applications in spin dependent electronics. This compound was used by several groups to produce thin films \cite{RRW01, GBW02, GBW03, KHM03, WPK05a, WPK05b} and devices \cite{IOT04, KTH04}. It was recently shown that Co$_2$FeSi is also a very good candidate as it exhibits a very high Curie temperature of 1100K and a magnetic moment of 6$\mu_B$ \cite{WFK05}. This compound is somewhat peculiar regarding its Curie temperature and magnetic moment because both are reaching the highest values observed up to now in this class of materials.

In Heusler as well as other compounds, the half-metallic ferromagnetism has not yet been unambiguously confirmed by experiments, although there is strong evidence in some cases. Despite a lot of work on Co and Mn containing Heusler compounds \cite{GBW02, FSI90, IMF98, RRH02, YYG03, HKS04, PCF04a, PCF04b, BCG04, DAX05, HKS05}, a breakthrough is still required to prove the existence of half metallic ferromagnetism in these materials. 

The electronic structure plays an important role in determining the magnetic properties of the Heusler compounds. In this work, a comprehensive investigation of the equilibrium structural, electronic and magnetic properties of Co$_2$FeSi is presented and compared to Co$_2$MnSi. The dependence of the magnetic moment on the lattice parameter was analyzed, focusing on the differences between LSDA (local spin density approximation) and GGA (generalized gradient approximation) treatments. A series of calculations for Co$_2$ based Heusler compounds was performed which shows that mostly all Co$_2$ based Heusler compounds exhibit half-metallic ferromagnetism \cite{FKW05a}, in agreement with Ref.\cite{GDP02}. The Slater-Pauling rule \cite{Sla36,Pau38,Kub84,GDP02,WFK05} fits for most Co$_2$ based Heusler compounds and may be used as a rule of thumb for estimating their magnetic moments. According to this rule the magnetic moment of Co$_2$FeSi is expected to be $6\mu_B$ per unit cell.

This work was stimulated by a recent experimental study that was devoted to the magnetic moment and Curie temperature of the Co$_2$FeSi compound \cite{WFK05}. To provide a concise picture of the Co$_2$YSi systems (Y = Mn, Fe), a theoretical investigation will be presented utilizing the local (spin) density approximation (L(S)DA), GGA, and LDA+$U$ methods. 

\section{Computational details}

Heusler compounds \cite{WZi88} belong to a group of ternary intermetallics with the stoichiometric composition X$_2$YZ ordered in an $L2_1$-type structure, space group $F\:m\overline{3}m$, many of which are ferromagnetic \cite{WZi73}. Remarkably, the prototype Cu$_2$MnAl is a ferromagnet even though none of its constituents is one \cite{Heu03}. In general, the X and Y atoms are transition metals and Z is a main group element. In some cases Y is replaced by a rare earth element. The X atoms are placed on 8a (1/4,1/4,1/4) Wyckoff positions and the Y and Z atoms on 4a (0,0,0) and 4b (1/2,1/2,1/2). The cubic $L2_1$ structure consists overall of four interpenetrating fcc lattices, two of them equally occupied by X. The latter two X$_2$ fcc sub-lattices combine to a simple cubic sub-lattice. Thus, the,  compound may be seen as a CsCl (B2) superstructure with the cubic vacancies of the X$_2$ sub-lattice filled alternatingly by Y or Z atoms. Other than in a simple CsCl structure, the nearest neighbor polyhedra of each $X$ atom are given by two different tetrahedra (rotated by $90^\circ$ with respect to each other) that are built from either Y or Z atoms. The $\Gamma$ point of the paramagnetic structure has the symmetry $O_h$. However, the wave functions at the $\Gamma$ point have to be described by $C_{4h}$ in the ferromagnetic state to account for the correct transformation of the electron spin.

Self consistent band structure calculations were carried out using the full potential linear augmented plane wave (FLAPW) method as provided by P. Blaha {\it et al} (Wien2k) \cite{BSM01}. The calculations were performed using the von~Barth-Hedin \cite{BHe72} parametrization of the exchange - correlation functional within LDA. Further, the Perdew-Burke-Ernzerhof \cite{PBE96} implementation of GGA was used. GGA accounts specifically for density gradients that are neglected in pure L(S)DA. The energy threshold between core and valence states was set to -6Ry. We considered 2.3a$_{Bohr}$ for the muffin-tin radii ($R_{MT}$) of all atoms leading to nearly touching spheres. 455 irreducible $k$ points of a $25\times25\times25$ mesh, were used for Brillouin zone integration. The number of plane waves was restricted by $R_{MT} \times k_{max} = 7$. The energy convergence criterion was set to $10^{-5}$. Simultaneously, the charge convergence was monitored and the calculation was restarted if it was still above $10^{-3}$. 

Finally, the LDA$+U$ method \cite{AAL97} was used to account for on-site correlation at the transition metal sites. The LDA$+U$ method accounts for an orbital dependence of the Coulomb and exchange interaction that is absent in pure LDA. Here, the effective Coulomb-exchange interaction ($U_{eff}=U-J$) was used for the calculations. This particular scheme is used in Wien2k to include double-counting correction. One leaves the framework of density functional theory (DFT) by applying the LDA$+U$ method. However, it is one of the most popular and easiest approaches to consider electron-electron correlation. It will be shown that the LDA$+U$ method gives qualitative and quantitative improvements compared to the bare LDA or GGA approaches, when applied to Co$_2$FeSi.

\section{Results and discussion}
 
The main focus was put on the magnetic moment as this shows a very strong discrepancy if comparing the experimental and calculated values reported for Co$_2$FeSi. Niculescu {\it et al} \cite{NBR77} reported already a magne, tic moment of 5.9$\mu_B$ per unit cell at 10K, whereas band structure calculations using the screened Korringa-Kohn-Rostocker method (KKR) predicted later only $5.27\mu_B$ \cite{GDP02}. The present LSDA calculations (FLAPW) revealed also a total magnetic moment of only 5.29$\mu_B$. This value is much lower than the measured value of about 6$\mu_B$ \cite{WFK05}. Such a large discrepancy points clearly that the actual electronic structure of this compound is different from the calculated one. Even so the calculations for Co$_2$MnSi agree with the experiment concerning the magnetic moment, their reliability concerning the details of the electronic structure may have to be called in question as long as the differences obtained for Co$_2$FeSi stay unexplained.

The discrepancy between experiment and calculation was not removed when using the GGA parametrization resulting still in a too low value ($5.56\mu_B$). Inspecting the spin resolved density of states (DOS) and band structure revealed the appearance of a gap in the minority bands, but located below the Fermi energy (see Figs.\ref{fig_8}b,\ref{fig_9}e in Sec.\ref{secios:ESLU}). This is finally one reason why the magnetic moment is too low and not an integer. In the following, details of the electronic structure calculations will be discussed.

\subsection{Structure optimization}
\label{secios:SOP}

First, a structural optimization was performed for Co$_2$FeSi and Co$_2$MnSi to determine whether the experimental lattice parameter minimizes the total energy. It was found that the optimized lattice parameter from the calculation agrees very well with experimental values of $a_{exp,Fe}=5.64$\AA and $a_{exp,Mn}=5.645$\AA. Within the FLAPW scheme the structure optimization for Co$_2$YSi (Y = Mn, Fe) was performed using the GGA parametrization and a spin polarized (ferromagnetic) setup. The initial crystal structural parameter for $\Delta a/a=0$ is taken from the experimental value $a_{exp}$ and varied in steps of 2\%. For each of the different $a$ parameter, the iteration procedure was performed up to self-consistency. The energy minimum then defines the optimal value of $a$.

\begin{figure}
\centering
\includegraphics[width=6cm]{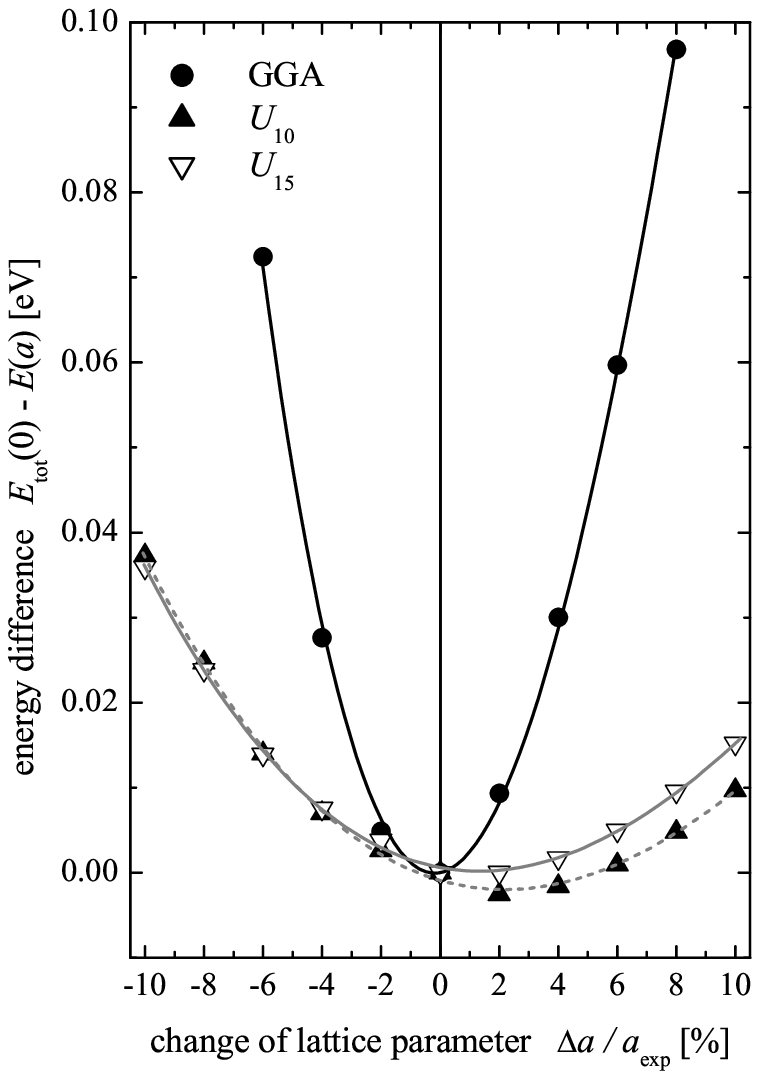}
\caption{Structural optimization for Co$_2$FeSi. \newline
         Shown is the change of the total energy as function of the lattice parameter ($a_0=5.64$\AA). 
         The GGA results are compared to LDA$+U$ calculations with different $U_{eff}$
         (see text). Note that all energy scales are shifted to result in $E_{tot}(0)=0$ for better 
         comparison. The lines are results from a polynomial fit.}
\label{fig_1}
\end{figure}

For Co$_2$FeSi, the energy minimum (ferromagnetic) was found to appear at $a= 5.63$\AA (corresponding to $\Delta a/a_{exp}\approx-0.2$\%, see Fig.\ref{fig_1}~GGA). It was also found that the ferromagnetic (spin polarized) configuration has the lower minimum total energy compared to the paramagnetic (non spin polarized) case, not shown here. Similar calculations revealed for Co$_2$MnSi that the energy minimum appears at $a= 5.651$\AA (corresponding to $\Delta a/a_{exp}\approx+0.1$\%.

Figure \ref{fig_1} displays in addition the results of a structural optimization using the LDA$+U$ scheme with $U_{eff}$ as variational parameter. The results of this optimization being performed for different values of $U_{eff}$ will be discussed below in Sec.\ref{secios:ECE}.

\subsubsection{Electronic structure and lattice parameter}
\label{secios:ESLP}

In order to understand why LSDA and GGA do not give the expected magnetic moment and position of the gap, the dependence of both on the lattice parameter was carefully inspected for Co$_2$FeSi as well as Co$_2$MnSi. It is important to compare the electronic structure of Co$_2$FeSi with Co$_2$MnSi as the latter is predicted to be a half-metallic ferromagnet with the measured magnetic moment of 5$\mu_B$ found in calculations using the experimental lattice parameter. However, a spin polarization of only about 55\% was obtained for this compound as determined by point contact Andreev reflection spectroscopy \cite{RXJ03, SBM04}. The spin polarization at the Fermi energy determined by means of photo emission was even lower ($8\ldots11$\%) \cite{WPK05a}. 

\begin{figure}
\centering
\includegraphics[width=8cm]{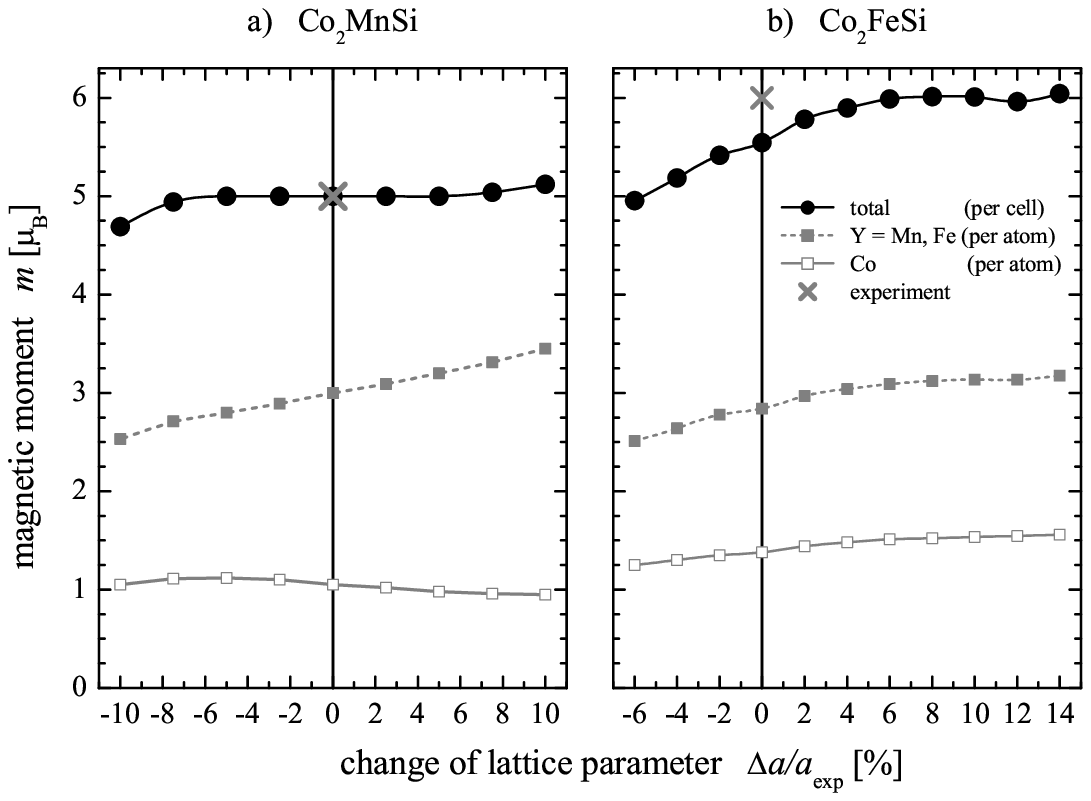}
\caption{Lattice parameter dependence of the magnetic moments. \newline
         Shown are the total and site specific magnetic moments of Co$_2$MnSi (a) and
         Co$_2$FeSi (b) as function of the lattice parameter.
         Experimental values are assigned by a cross.
         The experimental lattice parameters are $a_{Y=Mn}=5.645$\AA and $a_{Y=Fe}=5.64$\AA.
         Lines are drawn through the calculated values to guide the eye.}
\label{fig_2}
\end{figure}

The calculated total and site specific magnetic moments are shown in Fig.\ref{fig_2} for Co$_2$FeSi and Co$_2$MnSi. In Co$_2$FeSi (Fig.\ref{fig_2}b), the calculated atomic resolved magnetic moments of Co and Fe increase both with $a$ and the overall magnetic moment follows the same trend, it increases from $4.96\mu_B$ to above $6\mu_B$ with increase of $a$. The experimental value of $6\mu_B$ is found for an enlargement of the lattice parameter by about $6\ldots10$\%. The site specific moments seem to saturate above +6\% change of the lattice parameter at about 1.5$\mu_B$ and 3$\mu_B$ for Co and Fe, respectively. (Slightly higher values are compensated by an anti-parallel alignment of the moments at Si sites and in the interstitial.)

The total magnetic moment of Co$_2$MnSi (Fig.\ref{fig_2}a) increases also slightly with the lattice parameter but it stays at $5\mu_B$ in the range of $\pm6$\% change of $a$.  There is a significant change in the magnetic moment at Mn sites, it increases with lattice parameter. At the same time, the magnetic moment of Co decreases. Thus the Co moment counter balances the Mn moment such that the overall magnetic moment keeps constant while changing the lattice parameter.

At this point one can say that a moderate change of the lattice parameter does not change the overall magnetic moment of Co$_2$MnSi. The case of Co$_2$FeSi is completely different and an overall change of about 1$\mu_B$ is observed in the same range of $\Delta a/a$ as for Co$_2$MnSi. Overall, the magnetic moment of Co$_2$FeSi is less stable against variation of the lattice parameter (at least if using the same parameters for integration and convergence criteria as for the Co$_2$MnSi calculations). It exhibits some fluctuations about the integer value at very large lattice parameter.

The formation of the gap and localized magnetic moments in Heusler compounds is due to hybridization as was explained in detail by K\"ubler {\it et al}\cite{KWS83,Kub84}. There is a close relation between the magnetic moment and the HMF character. The appearance of the gap in the minority density constrains the number of minority electrons to be integer \footnote{This is true for ternary Heusler compounds, but may be different in quaternary compounds with non-integer site occupancies.}. However, an integer value of the magnetic moment may not result automatically in a real gap in the minority (or majority) density. The band structure of Co$_2$FeSi (see also below in Sec.\ref{secios:ESLU}) revealed already a small gap in the minority states in the calculation for the experimental lattice parameter, but this gap is located below the Fermi energy ($\epsilon_F$). Therefore, the band structure was closer inspected in order to proof whether the integer moment is related with the appearance of a real HMF minority gap.

\begin{figure}
\centering
\includegraphics[width=8cm]{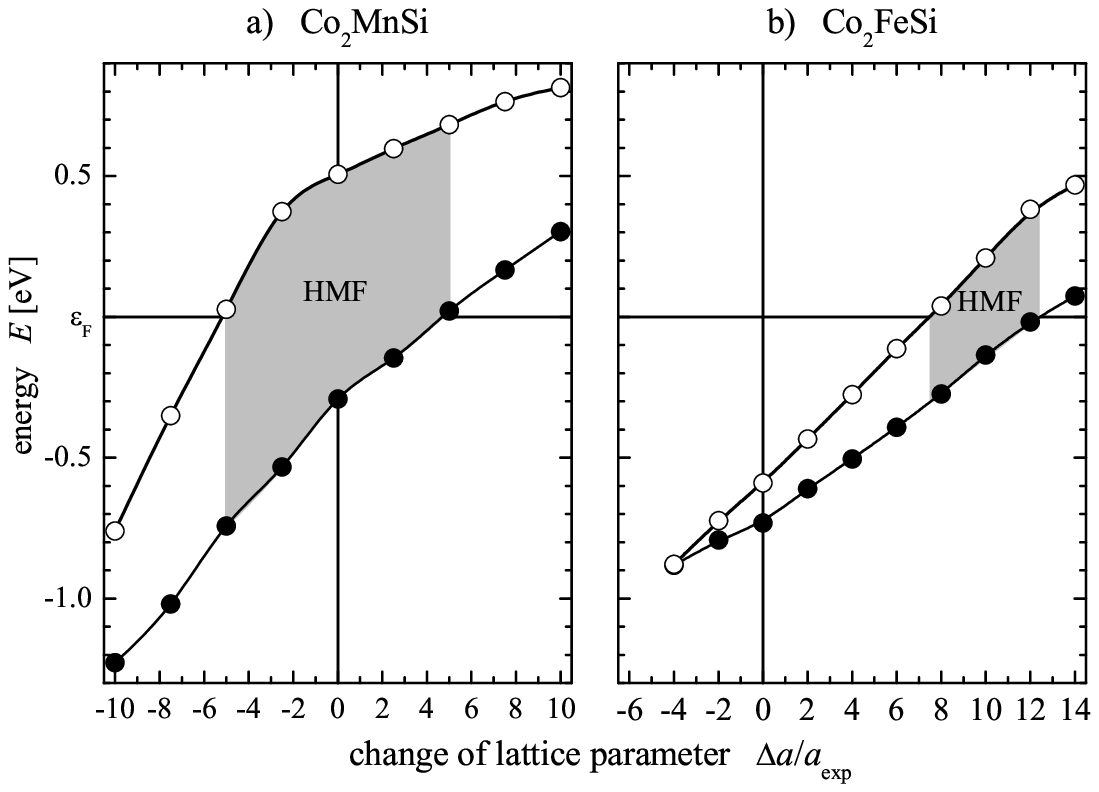}
\caption{Dependence of the minority band gap on the lattice parameter. \newline
         Shown are the extremal energies of the gap involving states for Co$_2$MnSi (a) and
         Co$_2$FeSi (b). The shaded areas assign the region of half-metallic ferromagnetism.
         Lines are drawn to guide the eye.}
\label{fig_3}
\end{figure}

Figure \ref{fig_3} shows the dependence of the extremal energies of the lower (valence) band and the upper (conduction) band of the minority states enveloping the gap. For both materials the magnetic moment has to be integer in the region were $\epsilon_F$ falls into the gap (gray shaded areas in Fig.\ref{fig_3}), that is the region of half-metallic ferromagnetism. 

Figure \ref{fig_3}a shows the dependence of the gap on the lattice parameter for Co$_2$MnSi. The shaded area corresponds to the region of half-metallic ferromagnetism. It is clearly seen that Co$_2$MnSi behaves like a HMF within $\pm5$\% change of the lattice parameter. Therefore, a moderate change of the lattice parameter does not change the HMF behavior of Co$_2$MnSi. Figure \ref{fig_3}b shows the dependence of the gap on the lattice parameter for Co$_2$FeSi. It is seen that the gap encloses the Fermi energy between 7.5\% and 12\% enlargement of the lattice parameter. The gap is completely closed for lattice parameter being 4\% smaller than $a_{exp}$. Obviously the gap ($\Delta E_{max}=0.4$eV at +12\%) is smaller compared to Co$_2$MnSi ($\Delta E_{max}=0.9$eV at -2.5\%) and appears in a narrower range of the lattice parameter. So far it is seen that the HMF character and thus the integer magnetic moment is more stable against variation of the lattice parameter in Co$_2$MnSi compared to Co$_2$FeSi.
 
An enlargement of the lattice parameter by $8\ldots12$\%, as needed to explain the magnetic moment of Co$_2$FeSi, corresponds to a volume-expansion of about $26\ldots40$\%. Such a large expansion of the crystal volume by about 1/3 is rather unrealistic and falls far out of the expected uncertainties for the experimental determination of $a$. However, it may be interesting to inspect which changes in the electronic structure are caused by such an expansion.

\begin{figure}
\centering
\includegraphics[width=8.5cm]{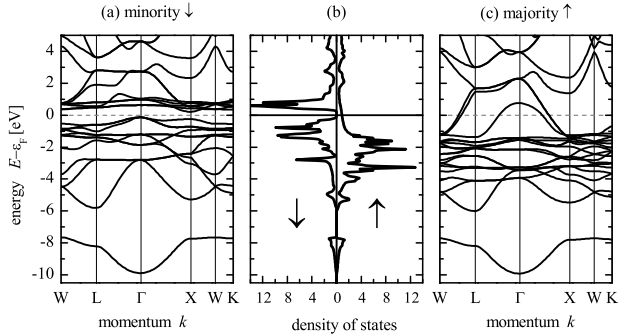}
\caption{Electronic structure of Co$_2$FeSi with enlarged lattice parameter. (LSDA-GGA, $a = 6.2$\AA )}
\label{fig_4}
\end{figure}

The density of states (DOS) and the band structure of Co$_2$FeSi are shown in Fig.\ref{fig_4} for the enlarged lattice parameter. The calculations were performed using the GGA scheme. An expansion of $\Delta a/a=10$\% was used in the calculation (for $\Delta a/a=$0 see Sec.\ref{secios:ESLU} and also Ref.\cite{WFK05}). This particular value was chosen as it is the case where the Fermi energy lies just in the middle of the gap of the minority DOS (see Fig.\ref{fig_4}a,b) and thus the material would securely be in a HMF state. 

There are no minority states at $\epsilon_F$ confirming the compound to be in a HMF state. The high density below $\epsilon_F$ is dominated by $d$ states being located at Co and Fe sites. The small density of states near $\epsilon_F$ in the majority DOS emerges from the strongly dispersing majority bands crossing the Fermi energy. From the spin resolved bands, it is seen that the majority bands cross or touch the Fermi energy ($\epsilon_F$) in rather all directions of high symmetry. On the other hand, the minority bands exhibit a clear gap around $\epsilon_F$. For Co$_2$FeSi, the width of the gap is given by the energies of the highest occupied band at the $\Gamma$-point and the lowest unoccupied band at the $\Gamma$ or $X$-point. The smaller value is found between $\Gamma$ and $X$, thus it is an indirect gap. The conclusion drawn from the displayed electronic structure is that the states around the Fermi energy are strongly spin polarized and the system is according to the calculations indeed a half-metallic ferromagnet, at least for an enlarged lattice parameter. The gap in the minority DOS has the compelling result that the magnetic moment is integer with a value of $6\mu_B$, as expected from the experiment. 

The change of the electronic structure with the lattice parameter is clear from two basic points of view. Firstly, from a free electron like band structure one expects a change of the band gradients and thus simply a shift of $\epsilon_F$ due to the accompanied changes in the DOS. Secondly, from a tight binding like approach one expects a difference in both, overlap and hopping integrals. In particular, they will be smaller for larger lattice parameter. With a decrease of $a$, the interaction between the atoms becomes stronger and the higher overlap results in a stronger de-localization of the electrons. Obviously the overlap is too large in the calculation using the experimental lattice parameter of Co$_2$FeSi. On the other hand one may make use for nearly HMF compounds that small deviations of the magnetic moment from integer values can be compensated by small variations of the lattice parameter supposed the LSDA (GGA) calculations result in the correct electronic structure. Unfortunately, the opposite is also true. If the calculated Fermi energy is not just in the middle of the minority gap, then a small deviation of the lattice parameter may destroy the HMF character.

The major conclusion that can be drawn from the needed increase of the lattice parameter of Co$_2$FeSi is that bare LSDA (GGA) calculations are not sufficient to explain its electronic and geometric structure at the same time and thus may fail for similar compounds, too. Moreover, it is suggested that electron-electron correlation might play an important role for opening the gap in the minority states and to get the magnetic moment with an integer value. The verification of the experimental results has to be fulfilled independent of the material for both, lattice parameter and magnetic moment. In the following section it will be analyzed whether this goal can be reached possibly by inclusion of correlation.

\subsection{Electron correlation}
\label{secios:ECE}

The relative importance of itinerant versus localized properties of $d$ electrons in metal alloys was already discussed by van Vleck \cite{VVl53} and Slater \cite{Sla36, Sla36a}. Heusler compounds are generally thought as systems exhibiting localized magnetic moments as was first mentioned by Pauling for Cu$_2$MnAl \cite{Pau38}. In particular, the magnetic moments of the Co$_2$YZ half-metallic ferromagnets follow strictly the {\it localized} part of the well known Slater-Pauling curve \cite{WFK05}. It is clear that the $d$ electrons are de-localized in metals. Thus, the question about correlation in transition metal based compounds is to what extent the on-site Coulomb interaction between $d$ electrons is preserved such that important atomic properties like Hund's rule correlation are significant and determine the magnetic properties at least partially \cite{Ful95}.

The idea from the above discussion is, after the structural optimization failed to explain the magnetic moment of Co$_2$FeSi, that inclusion of electron-electron correlation may be necessary in order to respect a partial {\it localization} of the $d$ electrons in a better way. Therefore, the LDA$+U$ method was used to recalculate the electronic structure of Co$_2$MnSi and Co$_2$FeSi.

\begin{figure}
\centering
\includegraphics[width=8cm]{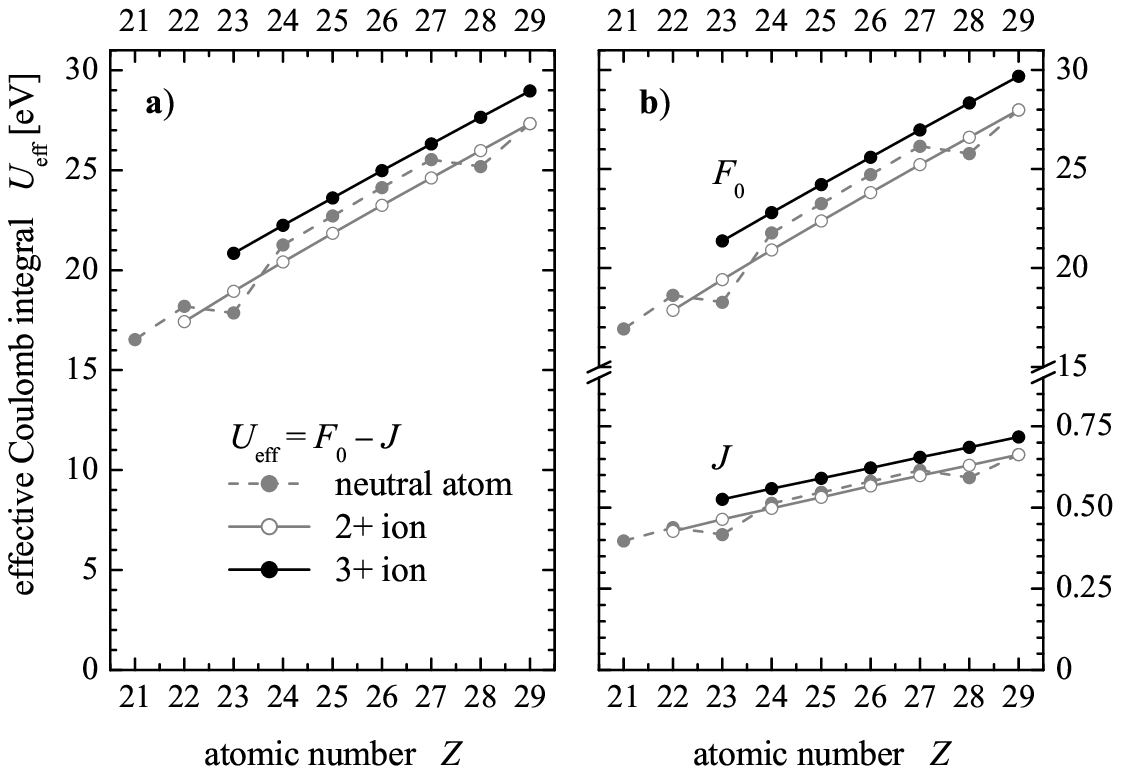}
\caption{Atomic Coulomb-exchange parameter. \newline
         The values were calculated for the $3d$ transition metals using the neutral as well as
         the most common ionic configurations.
         Note that the values for $J$ in (b) are shown on an enlarged scale below the break.} 
\label{fig_5}
\end{figure}

The LDA$+U$ scheme is designed to model localized states when on-site Coulomb interaction becomes important. It gives a self energy correction to localized states embedded in de-localized states. The energy $U$ of the Coulomb interaction is rather large in free atoms (17eV to 27eV in $3d$ transition metals, see Fig.\ref{fig_5}), while screening results in solids in a much smaller values \cite{SIm05} (for example 4.5eV in bcc Fe). In the present work, the value of the effective Coulomb-exchange interaction $U_{eff}=U-J$ was varied in order to reproduce the measured magnetic moment. It turned out that $U_{eff}$ from 2.5eV to 5.0eV for Co and simultaneously 2.4eV to 4.8eV for Fe result in a magnetic moment of $6\mu_B$. These values for $U_{eff}$ correspond to about $7\ldots20$\% of the free atom values as calculated from the corresponding Slater integrals by means of Cowan's program \cite{Cow81}. 

The average on-site Coulomb energy of pairs ($ij$) of equivalent electrons ($i=j$) is calculated from the Slater integrals $F_k$ to be:
\begin{equation}
E_{ii} = F_{0,ii} - \frac{2l_i+1}{4l_i+1} \sum 
         \left( \begin{array}{ccc} l_i \: k \: l_i \\ 0 \: 0 \: 0 \end{array} \right) ^2 F_{k,ii}
\label{eq0}
\end{equation}
The sum runs over all $k>0$ resulting in non-zero $3j$-symbols. The $3j$-symbols vanish for odd $k$ and for $d$ electrons one has $k_{max}=2l_i=4$ such that Eq.\ref{eq0} results in:
\begin{equation}
E_{dd} = F_0 - \frac{2}{63} F_2 - \frac{2}{63} F_4
\label{eq1}
\end{equation}
where the on-site Coulomb integral is $U=F_0$ and all integrals are calculated for $l=2$. The exchange integral is given by:
\begin{equation}
J = \frac{2}{63} (F_2 + F_4)
\label{eq2}
\end{equation}
Equation \ref{eq2} as calculated from Cowans's work \cite{Cow81} differs from the one reported by Anisimov {\it et al} \cite{AAL97} giving a factor 1/14 instead of 2/63, for unknown reasons \footnote{This effects also the exchange integrals for $f$ electrons, compare Ref.:\cite{Cow81} Tab.:~6-1 page 165 and Ref:\cite{AAL97}.}. 

The atomic values of $U=F_0$, $J$, and $U_{eff}=E_{dd}$ as calculated for the $3d$ transition metals are displayed in Fig.\ref{fig_5}. In the following we will denote the $U_{eff}$ values by $U_{x}$ where the subscript $x$ stands for the values (in \%) relative to the atomic ones (neutral atoms). It should be mentioned that the effect of screening in solids will mainly influence $U$ rather than the exchange integrals $J$. On the other hand, the values of $J$ depend also on the ionicity (see Fig.\ref{fig_5}b) that will be different in solids and dependent on the compound under investigation. From a practical point of view, the use of $U_{x}$ relative to the atomic values allows an easier comparison between different systems compared to the use of separate values for all quantities and atoms.

First, a structural optimization was performed for Co$_2$FeSi using different values of $U_{eff}$ in the calculations (see also Sec.\ref{secios:SOP}). From Fig.\ref{fig_1} it is seen that adding $U_{eff}$ results in a less pronounced change of the $E(a)$-dependence and shifts the energy minimum slightly to larger values of $a$. The minima are found at 5.72\AA and 5.75\AA for $U_{10}$ and $U_{15}$, respectively. These values correspond to $\Delta a/a_{exp}$ of about +1.6\% and thus are slightly higher compared to the experimental value. $U_{eff}$ has the effect that the total energies at the experimental lattice parameter are about 3eV lower for  $U_{15}$ compared to $U_{10}$. At the same lattice parameter, the total energy of the LSDA calculation is about 6 eV higher compared to $U_{10}$. It is worthwhile to note that the total energy taken from the LDA$+U$ calculation does not correspond to the LDA ground state energy anymore and thus may lead to wrong conclusions about the structure. However, $E(a,U_{eff})$ will still give an idea about the optimal structure.

\subsubsection{Magnetic moment and minority gap in LDA$+U$}
\label{secios:MMU}

In the following the dependence of the magnetic moments on the type of the exchange-correlation functional is analyzed and compared to the experimental values. The results of the calculations for Co$_2$FeSi using different approximations for the potential as well as the parametrization of the exchange-correlation part are summarized in Tab.\ref{tab_1}. 

\begin{table}
\centering
\caption{Magnetic moments of Co$_2$FeSi. \newline
         Given are the values calculated for $a=5.64$\AA using different calculation schemes (see text).
         All values are given in $\mu_B$. Total moments (m$_{tot}$) are given per unit
         cell and site resolved values (spin moment m$_s$, orbital moment m$_l$) are per atom. 
	 ($U$ was set to 15\% of the atomic values. +SO assigns calculations with spin-orbit interaction included. )}
\begin{ruledtabular}
\begin{tabular}{l|cc|cc|c}
              &  Co    &       &  Fe    &       & Co$_2$FeSi \\
              &  m$_s$ & m$_l$ &  m$_s$ & m$_l$ & m$_{tot}$  \\
\noalign{\smallskip}\hline                    
LSDA (vBH)    &  1.31  &       &  2.72  &       & 5.29 \\
LSDA (VWN)    &  1.40  &       &  2.87  &       & 5.59 \\
GGA (PBE)     &  1.39  &       &  2.85  &       & 5.56 \\
GGA + SO      &  1.38  & 0.04  &  2.83  & 0.06  & 5.56 \\
LDA$+U$       &  1.53  &       &  3.25  &       & 6.0  \\
LDA$+U$ + SO  &  1.56  & 0.08  &  3.24  & 0.07  & 6.0  \\
\end{tabular}
\end{ruledtabular}
\label{tab_1}     
\end{table}

Tabel \ref{tab_1} gives the site resolved moments at Co and Fe sites and the total magnetic moment. The induced moment at the Si sites (not given in Tab.\ref{tab_1} was aligned in all cases anti-parallel to that at the transition metal sites. Likewise, the magnetic moment located in the interstitial is omitted in Tab.\ref{tab_1}, however, both quantities have to be respected in order to find the correct total moment.

The lowest value for the total magnetic moment was found if using the v.~Barth-Hedin (vBH) \cite{BHe72} parametrization with a discrepancy of -0.7$\mu_B$ compared to the experiment. The results from Vosko-Wilk-Nussair \cite{VWN80} parametrization and GGA \cite{PBE96}  are very similar but still too low compared to the experiment. It is seen that both site specific moments are too low to reach the experimental value that may need values of about 1.5$\mu_B$ and 3$\mu_B$ per atom at the Co and Fe sites, respectively. From this point of view, the LSDA or GGA approaches are not sufficient to explain the magnetic structure of Co$_2$FeSi, independent of the type of parametrization. 

The correct magnetic moment at the experimental lattice parameter was only found if using the $+U$ functional. The use of the LDA$+U$ scheme improves the total magnetic moment considerably ($U_{15}$ was used for the calculation in Tab.\ref{tab_1}). The ratio of the magnetic moments of Co and Fe was measured to be $m_{Fe}/m_{Co}=2.2$ at 300K in an induction field of 0.4T \cite{WFK05}. The ratio of 2.1 found in the LDA$+U$ calculations agrees very well with this value. Spin-orbit interaction (SO) was additionally included in some of the calculations to analyze its influence on the total and partial magnetic moments. In the GGA calculations, it did not improve the total moment. The experiments delivered orbital to spin magnetic moment ratios ($m_{l}/m_{s}$) of about 0.05 for Fe and 0.1 for Co. The LDA$+U$+SO calculations revealed 0.02 for Fe and 0.05 for Co, being a factor of 2 smaller compared to the experiment. Overall, the agreement of both ratios ($m_{Fe}/m_{Co}$ and $m_{l}/m_{s}$) with the experiment is better in LDA$+U$ compared to the pure LSDA or GGA parametrization of the exchange-correlation functional. This may give advice that $U$ corrects, at least partially,  the missing orbital dependence of the potential in LSDA.

\begin{figure}
\centering
\includegraphics[width=8cm]{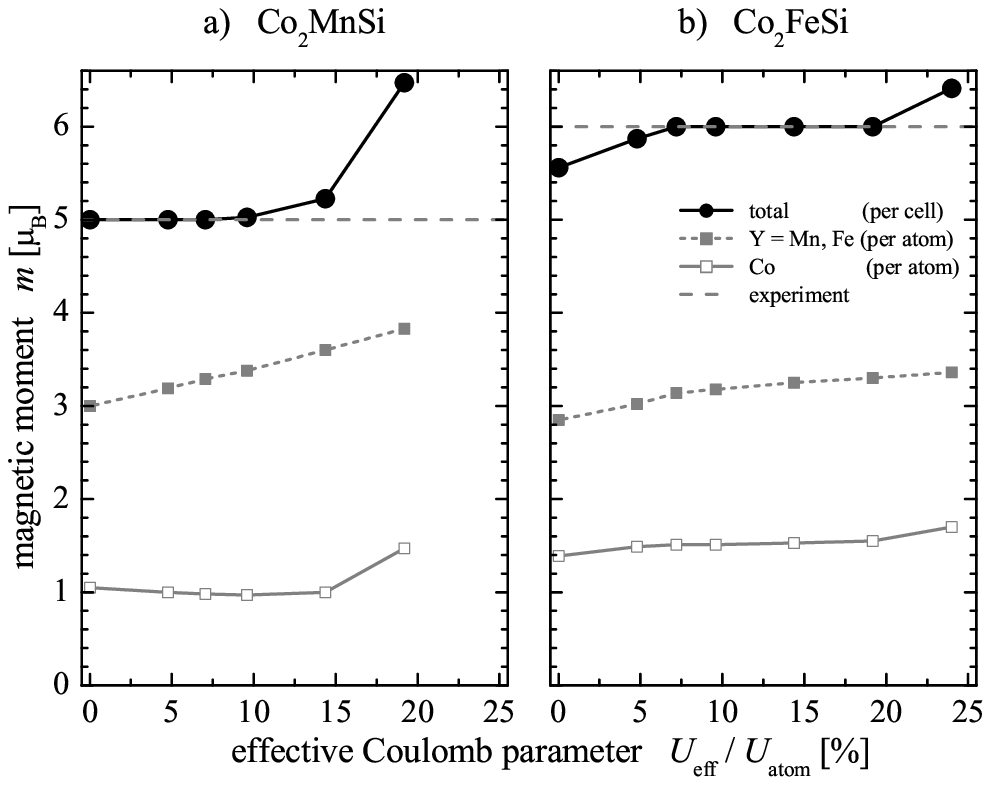}
\caption{Dependence of the magnetic moments on $U_{eff}$. \newline
         The dashed horizontal lines assign the experimental values of the magnetic moments for Co$_2$MnSi (a) and
         Co$_2$FeSi (b). The calculated values are connected by solid lines to guide the eye.} 
\label{fig_6}
\end{figure}

In the following the effect of the magnitude of the Coulomb-exchange interaction on the magnetic moments will be discussed. Figure \ref{fig_6} compares the dependence of the magnetic moment on $U_{eff}$ for Co$_2$FeSi and Co$_2$MnSi.

It is found for Co$_2$FeSi that the total magnetic moment increases from $5.6\mu_B$ to $6\mu_B$ while increasing $U_{eff}$ from 0 to $U_{7.5}$ and stays at the integer value up to $U_{20}$. Above that value the moment exhibits a further increase. The situation is different for Co$_2$MnSi with $5\mu_B$ at $U_{eff}=0$. The moment stays integer up to about $U_{7.5}$ and then increases with increasing $U_{eff}$.

In Co$_2$MnSi, the moment at the Mn sites increases nearly linearly with $U_{eff}$. At the same time, the moment decreases at the Co sites up to $U_{10}$ counter-balancing the Mn moments. At higher $U_{eff}$, the Co moment increases and the HMF character is lost. The situation is different in Co$_2$FeSi, both site specific moments show an overall slight increase with increasing $U_{eff}$. Moreover, they stay rather constant in the range of $U_{7.5}$ to $U_{20}$ such that the total moment stays integer in this region. In both materials, the moment at the Co sites reacts stronger on $U_{eff}$ than the ones at Y sites for $U_{eff}$ being above the region of integer total moments. The increase of the site specific moments is much less pronounced in Co$_2$FeSi compared to Co$_2$MnSi. This gives advice that the Fe compound is much more stable against correlation effects in comparison to the Mn compound, at least concerning the magnetic moments.  

\begin{figure}
\centering
\includegraphics[width=6cm]{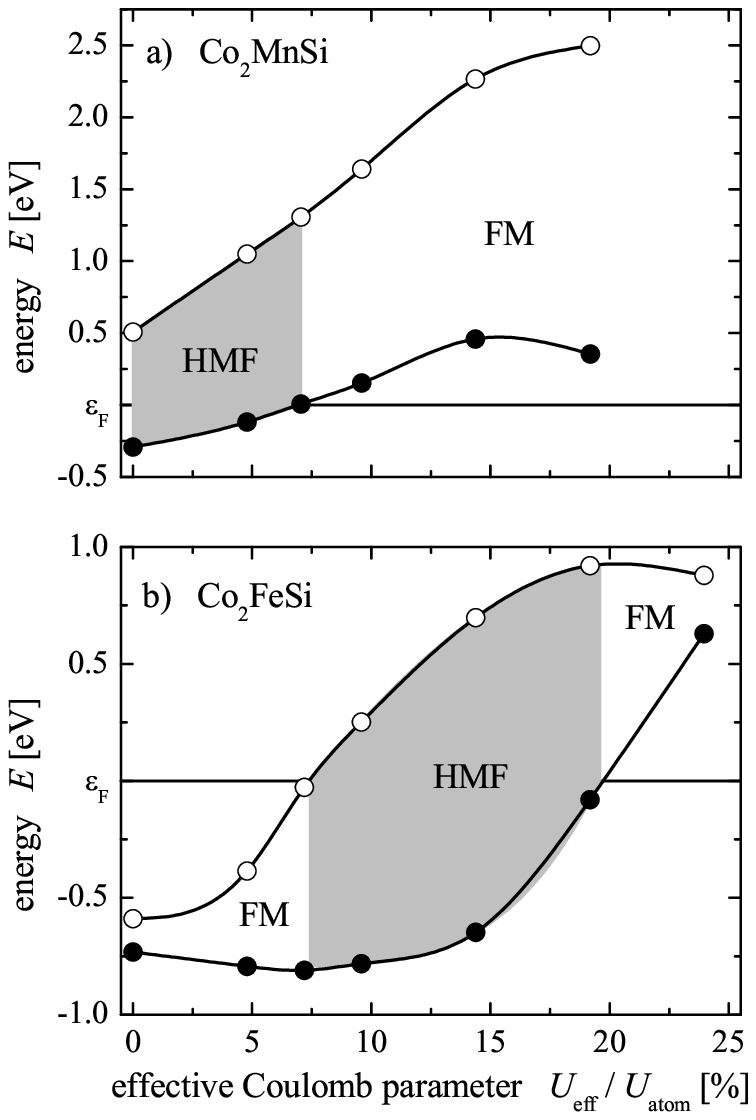}
\caption{Dependence of the minority gap on the effective Coulomb-exchange parameter. \newline
         Shown are the extremal energies of the gap involving states for Co$_2$MnSi (a) and
         Co$_2$FeSi (b). The shaded areas assign the region of half-metallic ferromagnetism.
         Lines are drawn to guide the eye.}
\label{fig_7}
\end{figure}

Again, it is interesting to find out whether the integer magnetic moment of Co$_2$FeSi is related to a real HMF gap in the minority band structure and to compare the behavior of the gap to the one of Co$_2$MnSi. The extremal energies of the gap enveloping bands are shown for both compounds in Fig.\ref{fig_7}.

In both cases the size of the gap increases with increasing $U_{eff}$. From Fig.\ref{fig_7}a it is found that Co$_2$MnSi stays in an HMF state in the range up to about $U_{8}$, that is the range of the integer magnetic moment. Taking larger values shifts $\epsilon_F$ to be outside the gap. In Co$_2$FeSi (Fig.\ref{fig_7}b), the minority gap involves $\epsilon_F$ from about $U_{8}$ up to $U_{20}$. That means, the integer value of the magnetic moment is related to the minority gap in both compounds and thus the direct consequence of the HMF state.

In both compounds, the gap is completely destroyed at very large values of $U_{eff}$ ($\geq 8$eV). The effect of the Coulomb-exchange interaction on single atoms (Co, Mn, or Fe) on the minority band gap was also investigated. It was found that the minority gap is destroyed in both materials if $U_{eff}$ is added only to one of the $3d$ elements, as expected. However, small changes in only one of the $U_{eff}$ values about the balanced value did not change the general behavior.

\subsubsection{Electronic structure in LDA$+U$}
\label{secios:ESLU}

\begin{figure}
\centering
\includegraphics[width=8cm]{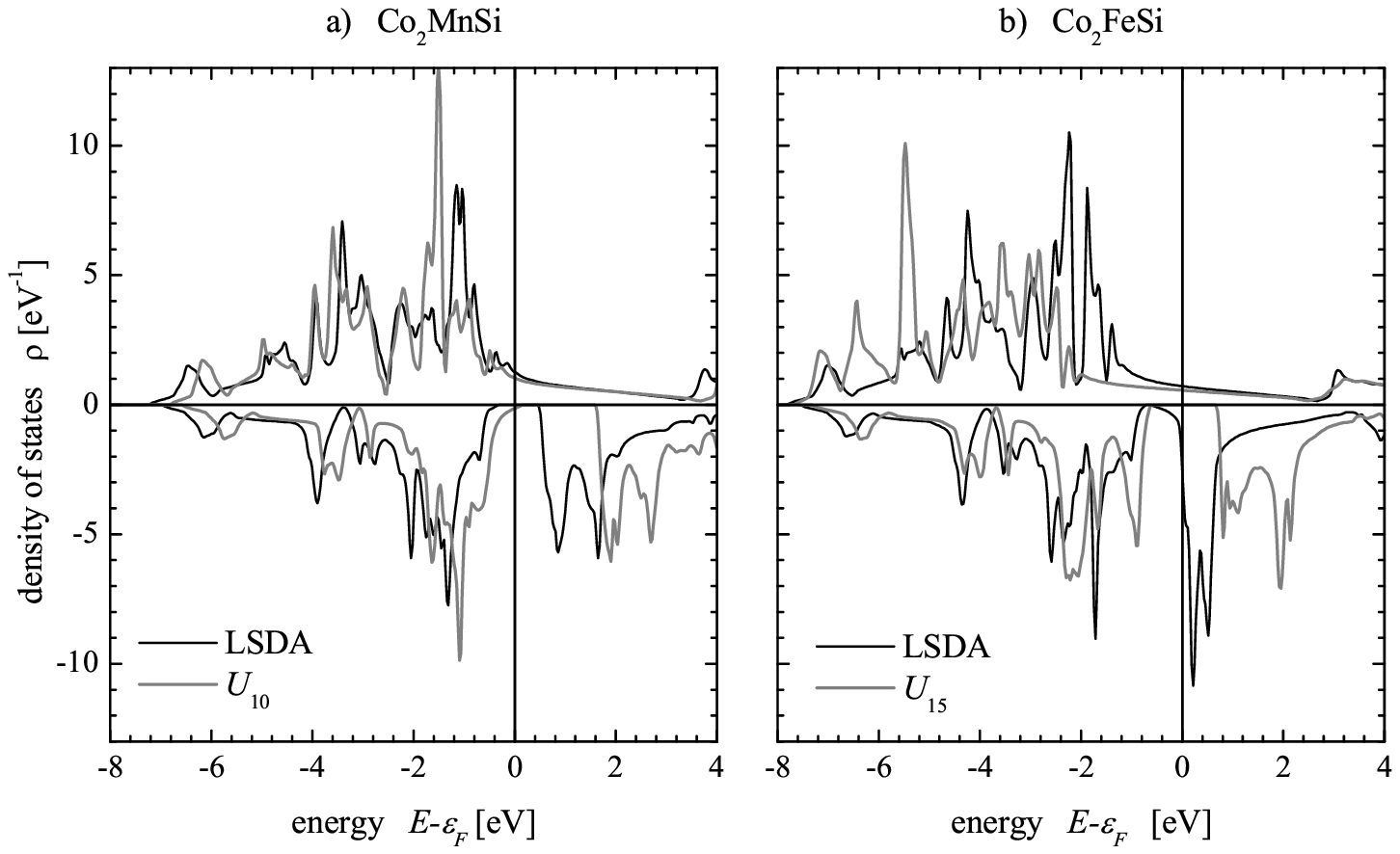}
\caption{Spin resolved densities of states for Co$_2$MnSi and Co$_2$FeSi. \newline
         Gray lines indicate the DOS within LDA$+U$ using $U_{10}$ for Co$_2$MnSi (a)
         and $U_{15}$ for Co$_2$FeSi (b). Black lines indicate the results from LSDA.
         The upper and lower parts of the plots display the majority and minority states, respectively.
	       (See text for the particular values of $U$ at the different sites.)}
\label{fig_8}
\end{figure}

In the following the influence of the correlation on the electronic structure will be discussed in more detail. Figure \ref{fig_8} compares the spin resolved density of states for Co$_2$MnSi and Co$_2$FeSi calculated in LSDA and LDA$+U$ approximation. The low lying $sp$-bands emerging from $\Gamma_{5g}$ (majority) and $\Gamma_{6g}$ (minority) $s$ states are located at $(-8\ldots-10)$eV. They are not shown here (compare Fig.\ref{fig_4}). These states are coupled to $p$ states ($\Gamma_{5u,6u}$) and mainly located at the Si atoms.

In both materials, the majority $d$ states spread from 8eV below to about 4eV above $\epsilon_F$ where the onset of the high lying, unoccupied $s$ states becomes visible. $s$ states are also present at below -4eV due to the $s-d$ coupling of the $\Gamma_{5g}$ states. $p$ states are contributing to the density over the hole range of the $d$ bands. They are mixed to $s$ and $d$ states due to the coupling of states with even and odd parity if going away from the $\Gamma$-point. The high density of states at -1eV in Co$_2$MnSi or -2eV in Co$_2$FeSi emerges mainly from $\Gamma_{8g}$ like states.
 
The situation is similar in the minority density of states where additionally a gap exists. This gap splits the minority $d$-states and the unoccupied part above $\epsilon_F$ is mainly of $\Gamma_{7g}$ character. The minority gap is clearly different in both compounds and depends strongly on the applied effective Coulomb-exchange interaction as already discussed above.

In the case of Co$_2$FeSi, where the experiment clearly indicates the total magnetic moment to have an integer value, the LSDA results seems to be incorrect. On the other hand, it should be noted that the unoccupied $d$ states are sharply peaked at $\epsilon_F$, resulting in a DOS that would imply an inherently unstable system in usual, simple ferromagnetic materials. This is despite the spin polarization. The system is, however, stabilized by the gap in the minority states that fixes the number of occupied states not only in the minority but also - and more important - in the majority channel. In LDA, the paramagnetic $e_g$ (located at the Co sites) and $t_{2g}$ (at Fe) states are rather narrow at $\epsilon_F$ suggesting that Coulomb correlation of the Hubbard U kind, being ignored within LSDA, will make Co$_2$FeSi insulating in the minority states. Attempting to treat electron correlation, the electrons become stronger localized by applying the LDA$+U$ method. One of the major issues here is that the LDA$+U$ type correlation is important for the electronic and magnetic structure of Co$_2$FeSi. In more detail, the majority states at around -4 eV (arising from $t_{2g}$ states in the paramagnetic case) are shifted by $U_{15}$ to -6eV below $\epsilon_F$. At the same time the unoccupied minority states seen in the LSDA calculation just above $\epsilon_F$ exhibit a larger shift of the paramagnetic $e_g$ states (to 2 eV above $\epsilon_F$) compared to the paramagnetic $t_{2g}$ states (to 0.5 eV above $\epsilon_F$). The occupied states of the minority DOS exhibit a less complicated behavior, they are mainly shifted towards the Fermi energy.

For Co$_2$FeSi as shown in Fig.\ref{fig_8}b, $U_{15}$ was chosen because for this value the Fermi energy lies just in the middle of the gap. It is clearly seen that the splitting between the occupied $\Gamma_{8g}$ and the unoccupied $\Gamma_{7g}$ states becomes larger. The minority $\Gamma_{7g}$ states exhibit an additional splitting after applying $+U$, whereas the unoccupied majority states stay rather unaffected. At the same time the gap in the minority states becomes considerably larger and the Fermi energy is clearly within this gap confirming that the compound is in a HMF state.

In Fig.\ref{fig_8}a the DOS of Co$_2$MnSi is compared in LSDA and LDA$+U$ approximation using $U_{10}$ being above the limit of the HMF state. The majority $d$ states being mainly located at the Co and Mn sites are clearly seen below $\epsilon_F$ and possess a width of about 7eV. Near $\epsilon_F$ one finds mainly a low density of majority states emerging from strongly dispersing $d$ bands. In LSDA, there are hardly any minority states at $\epsilon_F$ confirming the compound to be a half-metallic ferromagnet. Using LDA$+U$, the $d$ densities are shifting away from $\epsilon_F$ similar to Co$_2$FeSi. However, the occupied part of the minority DOS also changes if applying $U_{eff}$: it shifts closer to the Fermi energy and crosses it for high values of $U_{eff}$. Once it touches $\epsilon_F$ the maximum value of is reached that still results in the HMF state what is here the case for about $U_{7.5}$.

\begin{figure*}
\centering
\includegraphics[width=16cm]{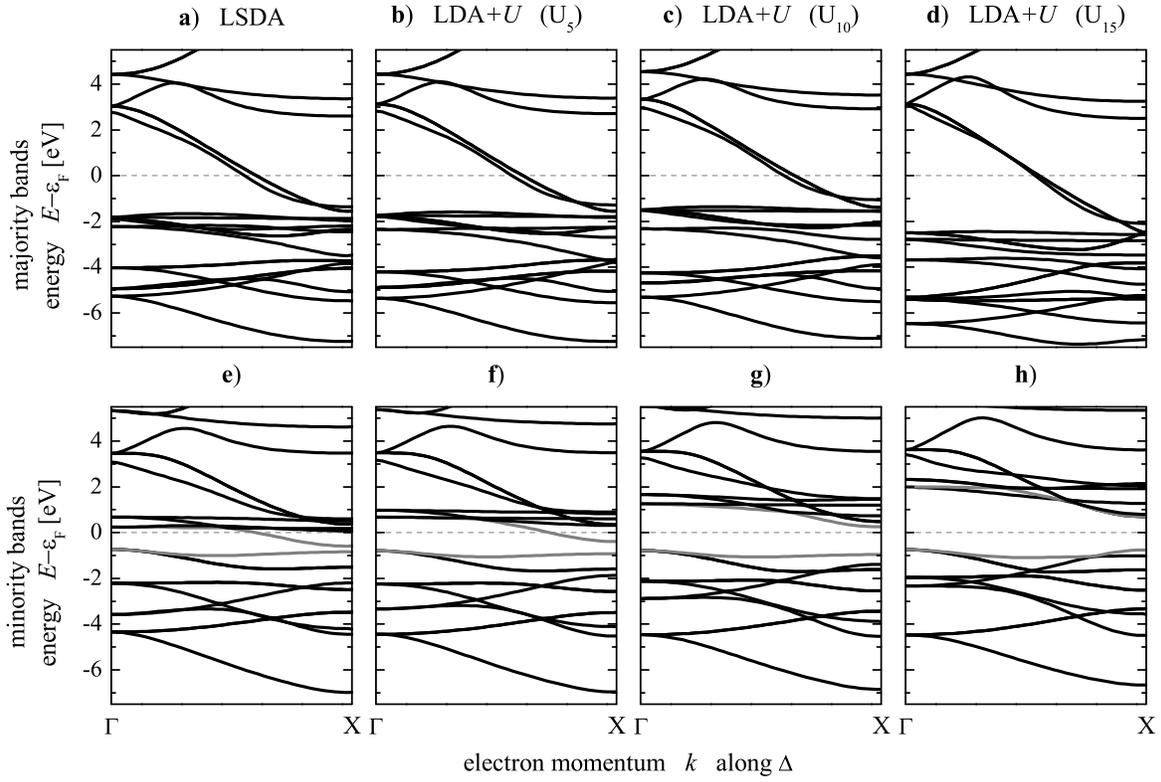}
\caption{Band structure of Co$_2$FeSi with variation of the Coulomb-exchange interaction.
         The gray lines assign the upper and lower bands defining the gap at the $X$-point in the minority states.
	       (See text for the particular values of $U_{\%}$ at the different sites.)} 
\label{fig_9}
\end{figure*}

Figure \ref{fig_9} shows the changes of the band structure of Co$_2$FeSi if applying the LDA$+U$ method. For easier comparison, details of the band structures are shown along $\Gamma-X$ that is the $\Delta$-direction of the paramagnetic state. The $\Delta$-direction is perpendicular to the Co$_2$ (100)-planes. As was shown earlier \cite{FKW05}, just the $\Delta$-direction plays the important role for the understanding of the HMF character and magnetic properties of Heusler compounds as was also pointed out by \"O\^g\"ut and Rabe \cite{ORa95}.

The developing of the band structure with increasing $U_{eff}$ is shown in Fig.\ref{fig_9} using values of 1.26eV, 2.52eV, and 3.78eV at the Co sites and simultaneously 1.2eV, 2.38eV, and 3.56eV at the Fe sites ($U_{5}$, $U_{10}$, and $U_{15}$). The upper panels (a-d) display majority and the lower panels (e-h) the minority band structures. It is seen that only slight changes appear in the majority band structure, the width of the visible bands at $\Gamma$ stays rather unaffected and the main changes appear close to $X$.

More interesting is the behavior of the minority bands as those determine the HMF character of the compound. One finds that the energies of the occupied states at $\Gamma$ stay nearly the same. The shapes of the bands close to $\Gamma$ are similar, too. The situation is different at $X$ where the unoccupied states are shifted away from $\epsilon_F$ resulting in a gap that increases with $U_{eff}$. It is clearly seen that it is an indirect gap. At $U_{10}$ the Fermi energy falls within the gap and the HMF state is reached. The minority gap is larger at $\Gamma$ or at other points of high symmetry (not shown here) compared to $X$. Therefore, it is obvious that the states at $X$ determine the HMF character of the compound.

The situation is clear at this stage: the gap in the minority states stays only within a certain range of the effective Coulomb-exchange interaction. If one goes away from that limit the Fermi energy falls not longer inside of the gap and the material loses its HMF character. It is also obvious that the mechanism leading to the half-metallic ferromagnetism in Co$_2$FeSi serves to destroy it in Co$_2$MnSi. The worst case would appear for both materials at about $7\ldots8$\% of the atomic values of $U_{eff}$ where LDA$+U$ predicts still values very close to the measured magnetic moments but also is the borderline for the loss of the HMF character. From this point of view, nearly integer magnetic moments alone do not verify the half-metallic ferromagnetism and one may have to search for alternative materials.

Previous band structure calculations using LSDA predicted Co$_2$MnSi to be a HMF at the experimental lattice parameter \cite{PCF02, GDP02}. Here, LDA$+U$ calculations were performed in order to consider the influence of on-site electron correlation in this compound. It was found that the gap in the minority DOS stays up to $U_{eff}=2.3$eV for Mn and 2.5eV for Co. For larger values of $U_{eff}$, that means for stronger correlation, the system loses its HMF character, by shifting $\epsilon_F$ outside the minority gap. Co$_2$MnSi is expected to exhibit 100\% spin polarization at the Fermi energy. In reality this was not verified up to now. Wang {\it et al} \cite{WPK05a} found only about 10\% spin polarization at room temperature from a spin-resolved photo emission experiment. The spectra shown in Ref.\cite{WPK05a} give no advice on the existence of a surface state that may serve to lower the spin polarization at $\epsilon_F$. Here, it was found that an already moderate correlation will destroy the gap. The spin polarization at $\epsilon_F$ amounts to 75\% if applying $U_{10}$, a value that is compatible to about 55\% found in Ref.:\cite{SBM04}. From that point of view, on-site correlation might be one more reason why no complete spin polarization was found in this compound (for others see: \cite{DSk04}).

Many effects have been considered to explain the missing experimental proof of the complete spin polarization in predicted half-metallic ferromagnets \cite{DSk04}. For sure, each of these effects may serve by itself to reduce the spin polarization at the Fermi energy. On the other hand, none of these effects is able to explain a difference of 15\% between the observed and calculated magnetic moments like for the case of Co$_2$FeSi. The appearance of surface or interface states \cite{WGr01, JKi02} may decrease the spin-polarization in certain layers but will not influence drastically the measurements of the magnetization of bulk samples. Magnon excitations \cite{JMT04, CVe02, BKJ01, HPC97, HRR00} will lead to a decrease of the magnetic moment with temperature rather than in an increase and thus can easily been disregarded here. Spin-orbit interaction couples, indeed, spin-up and spin-down states \cite{MSZ04} but does not enhance the magnetic moment, as was shown above. From this point of view, correlation might be one of the main causes of the too low spin polarization being observed in experiments. In the presented work, a static correction was used in the LDA$+U$ approximation. Dynamic correlation effects may be respected by means of the LDA$+DMFT$ approximation using dynamical mean field theory (DMFT). Using this method for NiMnSb, it was shown by Chioncel {\it et al}\cite{CKG03} that non quasi-particle effects may also serve to change the properties of the gap in the minority states. The use of bulk sensitive angle-, energy-, and spin- resolved photo emission becomes highly desirable in order to explain the details of the band structure in the most complete experimental way and to proof the existence of correlation in predicted half-metallic ferromagnets on the one hand and the true existence of half-metallic ferromagnetism on the other.

\section{Summary and Conclusions}

Structural parameters, magnetic moments and electronic properties of the Heusler compounds Co$_2$MnSi and Co$_2$FeSi were presented. The obtained, optimized lattice parameter leads for Co$_2$FeSi to a too small magnetic moment, using LSDA or GGA calculations. The measured value of the magnetic moment was only found for a lattice parameter enlarged by about 10\% compared to the experiment. It was also shown that the magnetic moment of Co$_2$MnSi is more stable against structural changes in the LSDA or GGA approaches. At present, it seems that simple LSDA and GGA methods are not sufficient to explain the electronic structure of Heusler compounds completely, at least they fail for Co$_2$FeSi.

In the step beyond LSDA, it was shown for the case of Co$_2$FeSi that electron correlation is able to explain the experimental magnetic moment at the experimentally observed lattice parameter. It was found that the LDA$+U$ scheme reproduces satisfactorily the experimental observations. At moderate effective Coulomb-exchange correlation energies of about 2.5eV to 4.5eV, the LDA$+U$ calculations agreed very well with the measured total and site specific magnetic moments. At the same time the compound was predicted to be clearly a half-metallic ferromagnet.

Finally, it was checked how the inclusion of correlation will change the calculated properties of similar Heusler compounds. Co$_2$MnSi is a compound being close to Co$_2$FeSi in the series of known Heusler compounds based on Co$_2$. It is obvious that on-site correlation will also play an important role in the Mn compound, if it does in the Fe compound. It was found that a small correlation energy of only about 2.5eV destroys the half-metallic ferromagnetism what may explain why a complete spin polarization was not observed in this compound up to now.

In conclusion, it is suggested that it will be necessary to include electron-electron correlation beyond LSDA and GGA in the theoretical description of all potential half-metallic ferromagnets and in particular for Heusler compounds.

\begin{acknowledgments}
This work is financially supported by the DFG (research project FG 559).
\end{acknowledgments}


\begin{thebibliography}{57}
\expandafter\ifx\csname natexlab\endcsname\relax\def\natexlab#1{#1}\fi
\expandafter\ifx\csname bibnamefont\endcsname\relax
  \def\bibnamefont#1{#1}\fi
\expandafter\ifx\csname bibfnamefont\endcsname\relax
  \def\bibfnamefont#1{#1}\fi
\expandafter\ifx\csname citenamefont\endcsname\relax
  \def\citenamefont#1{#1}\fi
\expandafter\ifx\csname url\endcsname\relax
  \def\url#1{\texttt{#1}}\fi
\expandafter\ifx\csname urlprefix\endcsname\relax\def\urlprefix{URL }\fi
\providecommand{\bibinfo}[2]{#2}
\providecommand{\eprint}[2][]{\url{#2}}

\bibitem[{\citenamefont{de~Groot et~al.}(1983)\citenamefont{de~Groot, Mueller,
  van Engen, and Buschow}}]{GME83}
\bibinfo{author}{\bibfnamefont{R.~A.} \bibnamefont{de~Groot}},
  \bibinfo{author}{\bibfnamefont{F.~M.} \bibnamefont{Mueller}},
  \bibinfo{author}{\bibfnamefont{P.~G.} \bibnamefont{van Engen}},
  \bibnamefont{and} \bibinfo{author}{\bibfnamefont{K.~H.~J.}
  \bibnamefont{Buschow}}, \bibinfo{journal}{Phys. Rev. Lett.}
  \textbf{\bibinfo{volume}{50}}, \bibinfo{pages}{2024} (\bibinfo{year}{1983}).

\bibitem[{\citenamefont{Ishida et~al.}(1995)\citenamefont{Ishida, Fujii,
  Kashiwagi, and Asano}}]{IFK95}
\bibinfo{author}{\bibfnamefont{S.}~\bibnamefont{Ishida}},
  \bibinfo{author}{\bibfnamefont{S.}~\bibnamefont{Fujii}},
  \bibinfo{author}{\bibfnamefont{S.}~\bibnamefont{Kashiwagi}},
  \bibnamefont{and} \bibinfo{author}{\bibfnamefont{S.}~\bibnamefont{Asano}},
  \bibinfo{journal}{J. Phys. Soc. Jap.} \textbf{\bibinfo{volume}{64}},
  \bibinfo{pages}{2152 } (\bibinfo{year}{1995}).

\bibitem[{\citenamefont{Brown et~al.}(2000)\citenamefont{Brown, Neumann,
  Webster, and Ziebeck}}]{BNW00}
\bibinfo{author}{\bibfnamefont{P.~J.} \bibnamefont{Brown}},
  \bibinfo{author}{\bibfnamefont{K.~U.} \bibnamefont{Neumann}},
  \bibinfo{author}{\bibfnamefont{P.~J.} \bibnamefont{Webster}},
  \bibnamefont{and} \bibinfo{author}{\bibfnamefont{K.~R.~A.}
  \bibnamefont{Ziebeck}}, \bibinfo{journal}{J. Phys. Condens Matter}
  \textbf{\bibinfo{volume}{12}}, \bibinfo{pages}{1827} (\bibinfo{year}{2000}).

\bibitem[{\citenamefont{Raphael et~al.}(2001)\citenamefont{Raphael, Ravel,
  Willard, Cheng, Das, Stroud, Bussmann, Claassen, and Harris}}]{RRW01}
\bibinfo{author}{\bibfnamefont{M.~P.} \bibnamefont{Raphael}},
  \bibinfo{author}{\bibfnamefont{B.}~\bibnamefont{Ravel}},
  \bibinfo{author}{\bibfnamefont{M.~A.} \bibnamefont{Willard}},
  \bibinfo{author}{\bibfnamefont{S.~F.} \bibnamefont{Cheng}},
  \bibinfo{author}{\bibfnamefont{B.~N.} \bibnamefont{Das}},
  \bibinfo{author}{\bibfnamefont{R.~M.} \bibnamefont{Stroud}},
  \bibinfo{author}{\bibfnamefont{K.~M.} \bibnamefont{Bussmann}},
  \bibinfo{author}{\bibfnamefont{J.~H.} \bibnamefont{Claassen}},
  \bibnamefont{and} \bibinfo{author}{\bibfnamefont{V.~G.}
  \bibnamefont{Harris}}, \bibinfo{journal}{Appl. Phys. Lett.}
  \textbf{\bibinfo{volume}{79}}, \bibinfo{pages}{4396 } (\bibinfo{year}{2001}).

\bibitem[{\citenamefont{Geiersbach et~al.}(2002)\citenamefont{Geiersbach,
  Bergmann, and Westerholt}}]{GBW02}
\bibinfo{author}{\bibfnamefont{U.}~\bibnamefont{Geiersbach}},
  \bibinfo{author}{\bibfnamefont{A.}~\bibnamefont{Bergmann}}, \bibnamefont{and}
  \bibinfo{author}{\bibfnamefont{K.}~\bibnamefont{Westerholt}},
  \bibinfo{journal}{J. Magn. Magn. Mater.} \textbf{\bibinfo{volume}{240}},
  \bibinfo{pages}{546 } (\bibinfo{year}{2002}).

\bibitem[{\citenamefont{Geiersbach et~al.}(2003)\citenamefont{Geiersbach,
  Bergmann, and Westerholt}}]{GBW03}
\bibinfo{author}{\bibfnamefont{U.}~\bibnamefont{Geiersbach}},
  \bibinfo{author}{\bibfnamefont{A.}~\bibnamefont{Bergmann}}, \bibnamefont{and}
  \bibinfo{author}{\bibfnamefont{K.}~\bibnamefont{Westerholt}},
  \bibinfo{journal}{Thin Solid Films} \textbf{\bibinfo{volume}{425}},
  \bibinfo{pages}{225 } (\bibinfo{year}{2003}).

\bibitem[{\citenamefont{K\"ammerer et~al.}(2003)\citenamefont{K\"ammerer,
  Heitmann, Meyners, Sudfeld, Thomas, H\"utten, and Reiss}}]{KHM03}
\bibinfo{author}{\bibfnamefont{S.}~\bibnamefont{K\"ammerer}},
  \bibinfo{author}{\bibfnamefont{S.}~\bibnamefont{Heitmann}},
  \bibinfo{author}{\bibfnamefont{D.}~\bibnamefont{Meyners}},
  \bibinfo{author}{\bibfnamefont{D.}~\bibnamefont{Sudfeld}},
  \bibinfo{author}{\bibfnamefont{A.}~\bibnamefont{Thomas}},
  \bibinfo{author}{\bibfnamefont{A.}~\bibnamefont{H\"utten}}, \bibnamefont{and}
  \bibinfo{author}{\bibfnamefont{G.}~\bibnamefont{Reiss}}, \bibinfo{journal}{J.
  Appl. Phys.} \textbf{\bibinfo{volume}{93}}, \bibinfo{pages}{7945}
  (\bibinfo{year}{2003}).

\bibitem[{\citenamefont{Wang et~al.}(2005{\natexlab{a}})\citenamefont{Wang,
  Przybylski, Kuch, Chelaru, Wang, Lu, Barthel, and Kirschner}}]{WPK05a}
\bibinfo{author}{\bibfnamefont{W.~H.} \bibnamefont{Wang}},
  \bibinfo{author}{\bibfnamefont{M.}~\bibnamefont{Przybylski}},
  \bibinfo{author}{\bibfnamefont{W.}~\bibnamefont{Kuch}},
  \bibinfo{author}{\bibfnamefont{L.~I.} \bibnamefont{Chelaru}},
  \bibinfo{author}{\bibfnamefont{J.}~\bibnamefont{Wang}},
  \bibinfo{author}{\bibfnamefont{Y.~F.} \bibnamefont{Lu}},
  \bibinfo{author}{\bibfnamefont{J.}~\bibnamefont{Barthel}}, \bibnamefont{and}
  \bibinfo{author}{\bibfnamefont{J.}~\bibnamefont{Kirschner}},
  \bibinfo{journal}{J. Magn. Magn. Mater.} \textbf{\bibinfo{volume}{286}},
  \bibinfo{pages}{336} (\bibinfo{year}{2005}{\natexlab{a}}).

\bibitem[{\citenamefont{Wang et~al.}(2005{\natexlab{b}})\citenamefont{Wang,
  Przybylski, Kuch, Chelaru, Wang, Lu, Barthel, Meyerheim, and
  Kirschner}}]{WPK05b}
\bibinfo{author}{\bibfnamefont{W.~H.} \bibnamefont{Wang}},
  \bibinfo{author}{\bibfnamefont{M.}~\bibnamefont{Przybylski}},
  \bibinfo{author}{\bibfnamefont{W.}~\bibnamefont{Kuch}},
  \bibinfo{author}{\bibfnamefont{L.~I.} \bibnamefont{Chelaru}},
  \bibinfo{author}{\bibfnamefont{J.}~\bibnamefont{Wang}},
  \bibinfo{author}{\bibfnamefont{F.}~\bibnamefont{Lu}},
  \bibinfo{author}{\bibfnamefont{J.}~\bibnamefont{Barthel}},
  \bibinfo{author}{\bibfnamefont{H.~L.} \bibnamefont{Meyerheim}},
  \bibnamefont{and}
  \bibinfo{author}{\bibfnamefont{J.}~\bibnamefont{Kirschner}},
  \bibinfo{journal}{Phys. Rev. B} \textbf{\bibinfo{volume}{71}},
  \bibinfo{pages}{144416} (\bibinfo{year}{2005}{\natexlab{b}}).

\bibitem[{\citenamefont{Inomata et~al.}(2004)\citenamefont{Inomata, Okamura,
  and Tezuka}}]{IOT04}
\bibinfo{author}{\bibfnamefont{K.}~\bibnamefont{Inomata}},
  \bibinfo{author}{\bibfnamefont{S.}~\bibnamefont{Okamura}}, \bibnamefont{and}
  \bibinfo{author}{\bibfnamefont{N.}~\bibnamefont{Tezuka}},
  \bibinfo{journal}{J. Magn. Magn. Mat.} \textbf{\bibinfo{volume}{282}},
  \bibinfo{pages}{269 } (\bibinfo{year}{2004}).

\bibitem[{\citenamefont{K\"ammerer et~al.}(2004)\citenamefont{K\"ammerer,
  Thomas, H\"utten, and Reiss}}]{KTH04}
\bibinfo{author}{\bibfnamefont{S.}~\bibnamefont{K\"ammerer}},
  \bibinfo{author}{\bibfnamefont{A.}~\bibnamefont{Thomas}},
  \bibinfo{author}{\bibfnamefont{A.}~\bibnamefont{H\"utten}}, \bibnamefont{and}
  \bibinfo{author}{\bibfnamefont{G.}~\bibnamefont{Reiss}},
  \bibinfo{journal}{Appl. Phys. Lett.} \textbf{\bibinfo{volume}{85}},
  \bibinfo{pages}{79} (\bibinfo{year}{2004}).

\bibitem[{\citenamefont{Wurmehl et~al.}(accepted)\citenamefont{Wurmehl, Fecher,
  Kandpal, Ksenofontov, and Felser}}]{WFK05}
\bibinfo{author}{\bibfnamefont{S.}~\bibnamefont{Wurmehl}},
  \bibinfo{author}{\bibfnamefont{G.~H.} \bibnamefont{Fecher}},
  \bibinfo{author}{\bibfnamefont{H.~C.} \bibnamefont{Kandpal}},
  \bibinfo{author}{\bibfnamefont{V.}~\bibnamefont{Ksenofontov}},
  \bibnamefont{and} \bibinfo{author}{\bibfnamefont{C.}~\bibnamefont{Felser}},
  \bibinfo{journal}{Phys. Rev. B} pp. \bibinfo{pages}{cond--mat / 0506729}
  (\bibinfo{year}{accepted}).

\bibitem[{\citenamefont{Fuji et~al.}(1990)\citenamefont{Fuji, Sugimura, Ishida,
  and Asano}}]{FSI90}
\bibinfo{author}{\bibfnamefont{S.}~\bibnamefont{Fuji}},
  \bibinfo{author}{\bibfnamefont{S.}~\bibnamefont{Sugimura}},
  \bibinfo{author}{\bibfnamefont{S.}~\bibnamefont{Ishida}}, \bibnamefont{and}
  \bibinfo{author}{\bibfnamefont{S.}~\bibnamefont{Asano}}, \bibinfo{journal}{J.
  Phys.: Condens. Matter} \textbf{\bibinfo{volume}{2}}, \bibinfo{pages}{8583 }
  (\bibinfo{year}{1990}).

\bibitem[{\citenamefont{Ishida et~al.}(1998)\citenamefont{Ishida, Masakai,
  Fujii, and Asano}}]{IMF98}
\bibinfo{author}{\bibfnamefont{S.}~\bibnamefont{Ishida}},
  \bibinfo{author}{\bibfnamefont{T.}~\bibnamefont{Masakai}},
  \bibinfo{author}{\bibfnamefont{S.}~\bibnamefont{Fujii}}, \bibnamefont{and}
  \bibinfo{author}{\bibfnamefont{S.}~\bibnamefont{Asano}},
  \bibinfo{journal}{Physica B} \textbf{\bibinfo{volume}{245}},
  \bibinfo{pages}{1 } (\bibinfo{year}{1998}).

\bibitem[{\citenamefont{Raphael et~al.}(2002)\citenamefont{Raphael, Ravel,
  Huang, Willard, Cheng, Das, Stroud, Bussmann, Claassen, and Harris}}]{RRH02}
\bibinfo{author}{\bibfnamefont{M.~P.} \bibnamefont{Raphael}},
  \bibinfo{author}{\bibfnamefont{B.}~\bibnamefont{Ravel}},
  \bibinfo{author}{\bibfnamefont{Q.}~\bibnamefont{Huang}},
  \bibinfo{author}{\bibfnamefont{M.~A.} \bibnamefont{Willard}},
  \bibinfo{author}{\bibfnamefont{S.~F.} \bibnamefont{Cheng}},
  \bibinfo{author}{\bibfnamefont{B.~N.} \bibnamefont{Das}},
  \bibinfo{author}{\bibfnamefont{R.~M.} \bibnamefont{Stroud}},
  \bibinfo{author}{\bibfnamefont{K.~M.} \bibnamefont{Bussmann}},
  \bibinfo{author}{\bibfnamefont{J.~H.} \bibnamefont{Claassen}},
  \bibnamefont{and} \bibinfo{author}{\bibfnamefont{V.~G.}
  \bibnamefont{Harris}}, \bibinfo{journal}{Phys. Rev. B}
  \textbf{\bibinfo{volume}{66}}, \bibinfo{pages}{104429}
  (\bibinfo{year}{2002}).

\bibitem[{\citenamefont{Yablonskikh et~al.}(2003)\citenamefont{Yablonskikh,
  Yarmoshenko, Gerasimov, Gaviko, Korotin, Kurmaev, Bartkowski, and
  Neumann}}]{YYG03}
\bibinfo{author}{\bibfnamefont{M.~V.} \bibnamefont{Yablonskikh}},
  \bibinfo{author}{\bibfnamefont{Y.~M.} \bibnamefont{Yarmoshenko}},
  \bibinfo{author}{\bibfnamefont{E.~G.} \bibnamefont{Gerasimov}},
  \bibinfo{author}{\bibfnamefont{V.~S.} \bibnamefont{Gaviko}},
  \bibinfo{author}{\bibfnamefont{M.~A.} \bibnamefont{Korotin}},
  \bibinfo{author}{\bibfnamefont{E.~J.} \bibnamefont{Kurmaev}},
  \bibinfo{author}{\bibfnamefont{S.}~\bibnamefont{Bartkowski}},
  \bibnamefont{and} \bibinfo{author}{\bibfnamefont{M.}~\bibnamefont{Neumann}},
  \bibinfo{journal}{J. Magn. Magn. Mater.} \textbf{\bibinfo{volume}{256}},
  \bibinfo{pages}{396} (\bibinfo{year}{2003}).

\bibitem[{\citenamefont{H\"utten et~al.}(2004)\citenamefont{H\"utten,
  K\"ammerer, Schmalhorst, and Thomas}}]{HKS04}
\bibinfo{author}{\bibfnamefont{A.}~\bibnamefont{H\"utten}},
  \bibinfo{author}{\bibfnamefont{S.}~\bibnamefont{K\"ammerer}},
  \bibinfo{author}{\bibfnamefont{J.}~\bibnamefont{Schmalhorst}},
  \bibnamefont{and} \bibinfo{author}{\bibfnamefont{A.}~\bibnamefont{Thomas}},
  \bibinfo{journal}{phys. stat. sol. (a)} \textbf{\bibinfo{volume}{201}},
  \bibinfo{pages}{3271 } (\bibinfo{year}{2004}).

\bibitem[{\citenamefont{Picozzi
  et~al.}(2004{\natexlab{a}})\citenamefont{Picozzi, Continenza, and
  Freeman}}]{PCF04a}
\bibinfo{author}{\bibfnamefont{S.}~\bibnamefont{Picozzi}},
  \bibinfo{author}{\bibfnamefont{A.}~\bibnamefont{Continenza}},
  \bibnamefont{and} \bibinfo{author}{\bibfnamefont{A.~J.}
  \bibnamefont{Freeman}}, \bibinfo{journal}{Phys. Rev. B}
  \textbf{\bibinfo{volume}{69}}, \bibinfo{pages}{094423}
  (\bibinfo{year}{2004}{\natexlab{a}}).

\bibitem[{\citenamefont{Picozzi
  et~al.}(2004{\natexlab{b}})\citenamefont{Picozzi, Continenza, and
  Freeman}}]{PCF04b}
\bibinfo{author}{\bibfnamefont{S.}~\bibnamefont{Picozzi}},
  \bibinfo{author}{\bibfnamefont{A.}~\bibnamefont{Continenza}},
  \bibnamefont{and} \bibinfo{author}{\bibfnamefont{A.~J.}
  \bibnamefont{Freeman}}, \bibinfo{journal}{J. Magn. Magn. Mat.}
  \textbf{\bibinfo{volume}{272-276}}, \bibinfo{pages}{315 }
  (\bibinfo{year}{2004}{\natexlab{b}}).

\bibitem[{\citenamefont{Block et~al.}(2004)\citenamefont{Block, Carey, Gurney,
  and Jepsen}}]{BCG04}
\bibinfo{author}{\bibfnamefont{T.}~\bibnamefont{Block}},
  \bibinfo{author}{\bibfnamefont{M.~J.} \bibnamefont{Carey}},
  \bibinfo{author}{\bibfnamefont{B.~A.} \bibnamefont{Gurney}},
  \bibnamefont{and} \bibinfo{author}{\bibfnamefont{O.}~\bibnamefont{Jepsen}},
  \bibinfo{journal}{Phys. Rev. B} \textbf{\bibinfo{volume}{70}},
  \bibinfo{pages}{205114} (\bibinfo{year}{2004}).

\bibitem[{\citenamefont{Dong et~al.}(2005)\citenamefont{Dong, Adelmann, Xie,
  Palmstrom, Lou, Strand, Crowell, Barnes, and Petford-Long}}]{DAX05}
\bibinfo{author}{\bibfnamefont{X.~Y.} \bibnamefont{Dong}},
  \bibinfo{author}{\bibfnamefont{C.}~\bibnamefont{Adelmann}},
  \bibinfo{author}{\bibfnamefont{J.~Q.} \bibnamefont{Xie}},
  \bibinfo{author}{\bibfnamefont{C.~J.} \bibnamefont{Palmstrom}},
  \bibinfo{author}{\bibfnamefont{X.}~\bibnamefont{Lou}},
  \bibinfo{author}{\bibfnamefont{J.}~\bibnamefont{Strand}},
  \bibinfo{author}{\bibfnamefont{P.~A.} \bibnamefont{Crowell}},
  \bibinfo{author}{\bibfnamefont{J.-P.} \bibnamefont{Barnes}},
  \bibnamefont{and} \bibinfo{author}{\bibfnamefont{A.~K.}
  \bibnamefont{Petford-Long}}, \bibinfo{journal}{Appl. Phys. Lett.}
  \textbf{\bibinfo{volume}{86}}, \bibinfo{pages}{102107}
  (\bibinfo{year}{2005}).

\bibitem[{\citenamefont{Hashemifar et~al.}(2005)\citenamefont{Hashemifar,
  Kratzer, and Scheffler}}]{HKS05}
\bibinfo{author}{\bibfnamefont{S.~J.} \bibnamefont{Hashemifar}},
  \bibinfo{author}{\bibfnamefont{P.}~\bibnamefont{Kratzer}}, \bibnamefont{and}
  \bibinfo{author}{\bibfnamefont{M.}~\bibnamefont{Scheffler}},
  \bibinfo{journal}{Phys. Rev. Lett.} \textbf{\bibinfo{volume}{94}},
  \bibinfo{pages}{096402} (\bibinfo{year}{2005}).

\bibitem[{\citenamefont{Fecher et~al.}(accepted)\citenamefont{Fecher, Kandpal,
  Wurmehl, Felser, and Sch\"onhense}}]{FKW05a}
\bibinfo{author}{\bibfnamefont{G.~H.} \bibnamefont{Fecher}},
  \bibinfo{author}{\bibfnamefont{H.~C.} \bibnamefont{Kandpal}},
  \bibinfo{author}{\bibfnamefont{S.}~\bibnamefont{Wurmehl}},
  \bibinfo{author}{\bibfnamefont{C.}~\bibnamefont{Felser}}, \bibnamefont{and}
  \bibinfo{author}{\bibfnamefont{G.}~\bibnamefont{Sch\"onhense}},
  \bibinfo{journal}{J. Appl. Phys.} pp. \bibinfo{pages}{cond--mat / 0510210}
  (\bibinfo{year}{accepted}).

\bibitem[{\citenamefont{Galanakis et~al.}(2002)\citenamefont{Galanakis,
  Dederichs, and Papanikolaou}}]{GDP02}
\bibinfo{author}{\bibfnamefont{I.}~\bibnamefont{Galanakis}},
  \bibinfo{author}{\bibfnamefont{P.~H.} \bibnamefont{Dederichs}},
  \bibnamefont{and}
  \bibinfo{author}{\bibfnamefont{N.}~\bibnamefont{Papanikolaou}},
  \bibinfo{journal}{Phys. Rev. B} \textbf{\bibinfo{volume}{66}},
  \bibinfo{pages}{174429} (\bibinfo{year}{2002}).

\bibitem[{\citenamefont{Slater}(1936{\natexlab{a}})}]{Sla36}
\bibinfo{author}{\bibfnamefont{J.~C.} \bibnamefont{Slater}},
  \bibinfo{journal}{Phys. Rev.} \textbf{\bibinfo{volume}{49}},
  \bibinfo{pages}{931} (\bibinfo{year}{1936}{\natexlab{a}}).

\bibitem[{\citenamefont{Pauling}(1938)}]{Pau38}
\bibinfo{author}{\bibfnamefont{L.}~\bibnamefont{Pauling}},
  \bibinfo{journal}{Phys. Rev.} \textbf{\bibinfo{volume}{54}},
  \bibinfo{pages}{899} (\bibinfo{year}{1938}).

\bibitem[{\citenamefont{K\"ubler}(1984)}]{Kub84}
\bibinfo{author}{\bibfnamefont{J.}~\bibnamefont{K\"ubler}},
  \bibinfo{journal}{Physica} \textbf{\bibinfo{volume}{127B}},
  \bibinfo{pages}{257 } (\bibinfo{year}{1984}).

\bibitem[{\citenamefont{Webster and Ziebeck}(1988)}]{WZi88}
\bibinfo{author}{\bibfnamefont{P.~J.} \bibnamefont{Webster}} \bibnamefont{and}
  \bibinfo{author}{\bibfnamefont{K.~R.~A.} \bibnamefont{Ziebeck}}, in
  \emph{\bibinfo{booktitle}{Alloys and Compounds of d-Elements with Main Group
  Elements. Part 2}}, edited by \bibinfo{editor}{\bibfnamefont{H.~P.~J.}
  \bibnamefont{Wijn}} (\bibinfo{publisher}{Springer-Verlag},
  \bibinfo{address}{Heidelberg}, \bibinfo{year}{1988}), vol.
  \bibinfo{volume}{19C} of \emph{\bibinfo{series}{Landolt-B\"ornstein - Group
  III Condensed Matter}}, pp. \bibinfo{pages}{104 -- 185}.

\bibitem[{\citenamefont{Webster and Ziebeck}(1973)}]{WZi73}
\bibinfo{author}{\bibfnamefont{P.~J.} \bibnamefont{Webster}} \bibnamefont{and}
  \bibinfo{author}{\bibfnamefont{K.~R.~A.} \bibnamefont{Ziebeck}},
  \bibinfo{journal}{J. Phys. Chem. Solids} \textbf{\bibinfo{volume}{34}},
  \bibinfo{pages}{1647} (\bibinfo{year}{1973}).

\bibitem[{\citenamefont{Heusler.}(1903)}]{Heu03}
\bibinfo{author}{\bibfnamefont{F.}~\bibnamefont{Heusler.}},
  \bibinfo{journal}{Verh. Dtsch. Phys. Ges.} \textbf{\bibinfo{volume}{12}},
  \bibinfo{pages}{219} (\bibinfo{year}{1903}).

\bibitem[{\citenamefont{Blaha et~al.}(2001)\citenamefont{Blaha, Schwarz,
  Madsen, Kvasnicka, and Luitz}}]{BSM01}
\bibinfo{author}{\bibfnamefont{P.}~\bibnamefont{Blaha}},
  \bibinfo{author}{\bibfnamefont{K.}~\bibnamefont{Schwarz}},
  \bibinfo{author}{\bibfnamefont{G.~K.~H.} \bibnamefont{Madsen}},
  \bibinfo{author}{\bibfnamefont{D.}~\bibnamefont{Kvasnicka}},
  \bibnamefont{and} \bibinfo{author}{\bibfnamefont{J.}~\bibnamefont{Luitz}},
  \emph{\bibinfo{title}{WIEN2k, An Augmented Plane Wave + Local Orbitals
  Program for Calculating Crystal Properties}} (\bibinfo{publisher}{Karlheinz
  Schwarz, Techn. Universitaet Wien}, \bibinfo{address}{Wien, Austria},
  \bibinfo{year}{2001}).

\bibitem[{\citenamefont{v.~Barth and Hedin}(1972)}]{BHe72}
\bibinfo{author}{\bibfnamefont{U.}~\bibnamefont{v.~Barth}} \bibnamefont{and}
  \bibinfo{author}{\bibfnamefont{L.}~\bibnamefont{Hedin}}, \bibinfo{journal}{J.
  Phys. C} \textbf{\bibinfo{volume}{5}}, \bibinfo{pages}{1629}
  (\bibinfo{year}{1972}).

\bibitem[{\citenamefont{Perdew et~al.}(1996)\citenamefont{Perdew, Burke, and
  Ernzerhof}}]{PBE96}
\bibinfo{author}{\bibfnamefont{J.~P.} \bibnamefont{Perdew}},
  \bibinfo{author}{\bibfnamefont{K.}~\bibnamefont{Burke}}, \bibnamefont{and}
  \bibinfo{author}{\bibfnamefont{M.}~\bibnamefont{Ernzerhof}},
  \bibinfo{journal}{Phys. Rev. Lett} \textbf{\bibinfo{volume}{77}},
  \bibinfo{pages}{3865} (\bibinfo{year}{1996}).

\bibitem[{\citenamefont{Anisimov et~al.}(1997)\citenamefont{Anisimov,
  Aryasetiawan, and Lichtenstein}}]{AAL97}
\bibinfo{author}{\bibfnamefont{V.~I.} \bibnamefont{Anisimov}},
  \bibinfo{author}{\bibfnamefont{F.}~\bibnamefont{Aryasetiawan}},
  \bibnamefont{and} \bibinfo{author}{\bibfnamefont{A.~I.}
  \bibnamefont{Lichtenstein}}, \bibinfo{journal}{J. Phys. Condens. Matter}
  \textbf{\bibinfo{volume}{9}}, \bibinfo{pages}{767} (\bibinfo{year}{1997}).

\bibitem[{\citenamefont{Niculescu et~al.}(1977)\citenamefont{Niculescu, Burch,
  Rai, and Budnick}}]{NBR77}
\bibinfo{author}{\bibfnamefont{V.}~\bibnamefont{Niculescu}},
  \bibinfo{author}{\bibfnamefont{T.~J.} \bibnamefont{Burch}},
  \bibinfo{author}{\bibfnamefont{K.}~\bibnamefont{Rai}}, \bibnamefont{and}
  \bibinfo{author}{\bibfnamefont{J.~I.} \bibnamefont{Budnick}},
  \bibinfo{journal}{J. Magn. Magn. Mater.} \textbf{\bibinfo{volume}{5}},
  \bibinfo{pages}{60} (\bibinfo{year}{1977}).

\bibitem[{\citenamefont{Ritchie et~al.}(2003)\citenamefont{Ritchie, Xiao, Ji,
  Chen, Chien, Zhang, Chen, Liu, Wu, and Zhang}}]{RXJ03}
\bibinfo{author}{\bibfnamefont{L.}~\bibnamefont{Ritchie}},
  \bibinfo{author}{\bibfnamefont{G.}~\bibnamefont{Xiao}},
  \bibinfo{author}{\bibfnamefont{Y.}~\bibnamefont{Ji}},
  \bibinfo{author}{\bibfnamefont{T.~Y.} \bibnamefont{Chen}},
  \bibinfo{author}{\bibfnamefont{C.~L.} \bibnamefont{Chien}},
  \bibinfo{author}{\bibfnamefont{M.}~\bibnamefont{Zhang}},
  \bibinfo{author}{\bibfnamefont{J.}~\bibnamefont{Chen}},
  \bibinfo{author}{\bibfnamefont{Z.}~\bibnamefont{Liu}},
  \bibinfo{author}{\bibfnamefont{G.}~\bibnamefont{Wu}}, \bibnamefont{and}
  \bibinfo{author}{\bibfnamefont{X.~X.} \bibnamefont{Zhang}},
  \bibinfo{journal}{Phys. Rev. B} \textbf{\bibinfo{volume}{68}},
  \bibinfo{pages}{104330} (\bibinfo{year}{2003}).

\bibitem[{\citenamefont{Singh et~al.}(2004)\citenamefont{Singh, Barber,
  Miyoshi, Bugoslavsky, Branford, and Cohen}}]{SBM04}
\bibinfo{author}{\bibfnamefont{L.~J.} \bibnamefont{Singh}},
  \bibinfo{author}{\bibfnamefont{Z.~H.} \bibnamefont{Barber}},
  \bibinfo{author}{\bibfnamefont{Y.}~\bibnamefont{Miyoshi}},
  \bibinfo{author}{\bibfnamefont{Y.}~\bibnamefont{Bugoslavsky}},
  \bibinfo{author}{\bibfnamefont{W.~R.} \bibnamefont{Branford}},
  \bibnamefont{and} \bibinfo{author}{\bibfnamefont{L.~F.} \bibnamefont{Cohen}},
  \bibinfo{journal}{Appl. Phys. Lett.} \textbf{\bibinfo{volume}{84}},
  \bibinfo{pages}{2367} (\bibinfo{year}{2004}).

\bibitem[{\citenamefont{J.~K\"ubler et~al.}(1983)\citenamefont{J.~K\"ubler,
  Williams, and Sommers}}]{KWS83}
\bibinfo{author}{\bibfnamefont{J.}~\bibnamefont{J.~K\"ubler}},
  \bibinfo{author}{\bibfnamefont{A.~R.} \bibnamefont{Williams}},
  \bibnamefont{and} \bibinfo{author}{\bibfnamefont{C.~B.}
  \bibnamefont{Sommers}}, \bibinfo{journal}{Phys. Rev. B}
  \textbf{\bibinfo{volume}{28}}, \bibinfo{pages}{1745} (\bibinfo{year}{1983}).

\bibitem[{\citenamefont{van Vleck}(1953)}]{VVl53}
\bibinfo{author}{\bibfnamefont{J.~H.} \bibnamefont{van Vleck}},
  \bibinfo{journal}{Rev. Mod. Phys.} \textbf{\bibinfo{volume}{25}},
  \bibinfo{pages}{220} (\bibinfo{year}{1953}).

\bibitem[{\citenamefont{Slater}(1936{\natexlab{b}})}]{Sla36a}
\bibinfo{author}{\bibfnamefont{J.~C.} \bibnamefont{Slater}},
  \bibinfo{journal}{Phys. Rev. B} \textbf{\bibinfo{volume}{49}},
  \bibinfo{pages}{537} (\bibinfo{year}{1936}{\natexlab{b}}).

\bibitem[{\citenamefont{Fulde}(1995)}]{Ful95}
\bibinfo{author}{\bibfnamefont{P.}~\bibnamefont{Fulde}},
  \emph{\bibinfo{title}{Electron correlations in molecules and solids, 3. ed.}}
  (\bibinfo{publisher}{Springer-Verlag}, \bibinfo{address}{Heidelberg},
  \bibinfo{year}{1995}).

\bibitem[{\citenamefont{Solovyev and Imada}(2005)}]{SIm05}
\bibinfo{author}{\bibfnamefont{I.~V.} \bibnamefont{Solovyev}} \bibnamefont{and}
  \bibinfo{author}{\bibfnamefont{M.}~\bibnamefont{Imada}},
  \bibinfo{journal}{Phys. Rev. B} \textbf{\bibinfo{volume}{71}},
  \bibinfo{pages}{045103} (\bibinfo{year}{2005}).

\bibitem[{\citenamefont{Cowan}(1981)}]{Cow81}
\bibinfo{author}{\bibfnamefont{R.~D.} \bibnamefont{Cowan}},
  \emph{\bibinfo{title}{The Theory of Atomic Structure and Spectra}}
  (\bibinfo{publisher}{University of California Press},
  \bibinfo{address}{Berkeley and Los Angeles}, \bibinfo{year}{1981}).

\bibitem[{\citenamefont{Vosko et~al.}(1980)\citenamefont{Vosko, Wilk, and
  Nussair}}]{VWN80}
\bibinfo{author}{\bibfnamefont{S.~H.} \bibnamefont{Vosko}},
  \bibinfo{author}{\bibfnamefont{L.}~\bibnamefont{Wilk}}, \bibnamefont{and}
  \bibinfo{author}{\bibfnamefont{M.}~\bibnamefont{Nussair}},
  \bibinfo{journal}{Can. J. Phys} \textbf{\bibinfo{volume}{58}},
  \bibinfo{pages}{1200} (\bibinfo{year}{1980}).

\bibitem[{\citenamefont{Fecher et~al.}(2005)\citenamefont{Fecher, Kandpal,
  Wurmehl, Morais, Lin, Elemrs, Sch\"onhense, and Felser}}]{FKW05}
\bibinfo{author}{\bibfnamefont{G.~H.} \bibnamefont{Fecher}},
  \bibinfo{author}{\bibfnamefont{H.~C.} \bibnamefont{Kandpal}},
  \bibinfo{author}{\bibfnamefont{S.}~\bibnamefont{Wurmehl}},
  \bibinfo{author}{\bibfnamefont{J.}~\bibnamefont{Morais}},
  \bibinfo{author}{\bibfnamefont{H.-J.} \bibnamefont{Lin}},
  \bibinfo{author}{\bibfnamefont{H.-J.} \bibnamefont{Elemrs}},
  \bibinfo{author}{\bibfnamefont{G.}~\bibnamefont{Sch\"onhense}},
  \bibnamefont{and} \bibinfo{author}{\bibfnamefont{C.}~\bibnamefont{Felser}},
  \bibinfo{journal}{J. Phys. Condens. Matter}
  \textbf{\bibinfo{volume}{17(46)}}, \bibinfo{pages}{7237}
  (\bibinfo{year}{2005}).

\bibitem[{\citenamefont{\"O\^g\"ut and Rabe}(1995)}]{ORa95}
\bibinfo{author}{\bibfnamefont{S.}~\bibnamefont{\"O\^g\"ut}} \bibnamefont{and}
  \bibinfo{author}{\bibfnamefont{K.~M.} \bibnamefont{Rabe}},
  \bibinfo{journal}{Phys. Rev. B} \textbf{\bibinfo{volume}{51}},
  \bibinfo{pages}{10443 } (\bibinfo{year}{1995}).

\bibitem[{\citenamefont{Picozzi et~al.}(2002)\citenamefont{Picozzi, Continenza,
  and Freeman}}]{PCF02}
\bibinfo{author}{\bibfnamefont{S.}~\bibnamefont{Picozzi}},
  \bibinfo{author}{\bibfnamefont{A.}~\bibnamefont{Continenza}},
  \bibnamefont{and} \bibinfo{author}{\bibfnamefont{A.~J.}
  \bibnamefont{Freeman}}, \bibinfo{journal}{Phys. Rev. B}
  \textbf{\bibinfo{volume}{66}}, \bibinfo{pages}{094421}
  (\bibinfo{year}{2002}).

\bibitem[{\citenamefont{Dowben and Skomski}(2004)}]{DSk04}
\bibinfo{author}{\bibfnamefont{P.~A.} \bibnamefont{Dowben}} \bibnamefont{and}
  \bibinfo{author}{\bibfnamefont{R.}~\bibnamefont{Skomski}},
  \bibinfo{journal}{J. Appl. Phys.} \textbf{\bibinfo{volume}{95}},
  \bibinfo{pages}{7453} (\bibinfo{year}{2004}).

\bibitem[{\citenamefont{de~Wijs and de~Groot}(2001)}]{WGr01}
\bibinfo{author}{\bibfnamefont{G.~A.} \bibnamefont{de~Wijs}} \bibnamefont{and}
  \bibinfo{author}{\bibfnamefont{R.~A.} \bibnamefont{de~Groot}},
  \bibinfo{journal}{Phys. Rev. B} \textbf{\bibinfo{volume}{64}},
  \bibinfo{pages}{020402} (\bibinfo{year}{2001}).

\bibitem[{\citenamefont{Jenkins and King}(2002)}]{JKi02}
\bibinfo{author}{\bibfnamefont{S.~J.} \bibnamefont{Jenkins}} \bibnamefont{and}
  \bibinfo{author}{\bibfnamefont{D.~A.} \bibnamefont{King}},
  \bibinfo{journal}{Surf. Sci. Lett.} \textbf{\bibinfo{volume}{501}},
  \bibinfo{pages}{L185} (\bibinfo{year}{2002}).

\bibitem[{\citenamefont{Jo et~al.}(2000)\citenamefont{Jo, Mathur, Todd, and
  Blamire}}]{JMT04}
\bibinfo{author}{\bibfnamefont{M.~H.} \bibnamefont{Jo}},
  \bibinfo{author}{\bibfnamefont{N.~D.} \bibnamefont{Mathur}},
  \bibinfo{author}{\bibfnamefont{N.~K.} \bibnamefont{Todd}}, \bibnamefont{and}
  \bibinfo{author}{\bibfnamefont{M.~G.} \bibnamefont{Blamire}},
  \bibinfo{journal}{Phys. Rev. B} \textbf{\bibinfo{volume}{61}},
  \bibinfo{pages}{14905} (\bibinfo{year}{2000}).

\bibitem[{\citenamefont{Coey and Venkatesan}(2002)}]{CVe02}
\bibinfo{author}{\bibfnamefont{J.~M.~D.} \bibnamefont{Coey}} \bibnamefont{and}
  \bibinfo{author}{\bibfnamefont{M.}~\bibnamefont{Venkatesan}},
  \bibinfo{journal}{J. Appl. Phys.} \textbf{\bibinfo{volume}{91}},
  \bibinfo{pages}{8345} (\bibinfo{year}{2002}).

\bibitem[{\citenamefont{Borca et~al.}(2001)\citenamefont{Borca, Komesu, Jeong,
  Dowben, Ristoiu, Hordequin, Nozieres, Pierre, Stadler, and Idzerda}}]{BKJ01}
\bibinfo{author}{\bibfnamefont{C.~N.} \bibnamefont{Borca}},
  \bibinfo{author}{\bibfnamefont{T.}~\bibnamefont{Komesu}},
  \bibinfo{author}{\bibfnamefont{H.-K.} \bibnamefont{Jeong}},
  \bibinfo{author}{\bibfnamefont{P.~A.} \bibnamefont{Dowben}},
  \bibinfo{author}{\bibfnamefont{D.}~\bibnamefont{Ristoiu}},
  \bibinfo{author}{\bibfnamefont{C.}~\bibnamefont{Hordequin}},
  \bibinfo{author}{\bibfnamefont{J.~P.} \bibnamefont{Nozieres}},
  \bibinfo{author}{\bibfnamefont{J.}~\bibnamefont{Pierre}},
  \bibinfo{author}{\bibfnamefont{S.}~\bibnamefont{Stadler}}, \bibnamefont{and}
  \bibinfo{author}{\bibfnamefont{Y.~U.} \bibnamefont{Idzerda}},
  \bibinfo{journal}{Phys. Rev. B} \textbf{\bibinfo{volume}{64}},
  \bibinfo{pages}{052409} (\bibinfo{year}{2001}).

\bibitem[{\citenamefont{Hordequin et~al.}(1997)\citenamefont{Hordequin, Pierre,
  and Currat}}]{HPC97}
\bibinfo{author}{\bibfnamefont{C.}~\bibnamefont{Hordequin}},
  \bibinfo{author}{\bibfnamefont{J.}~\bibnamefont{Pierre}}, \bibnamefont{and}
  \bibinfo{author}{\bibfnamefont{R.}~\bibnamefont{Currat}},
  \bibinfo{journal}{Physica B} \textbf{\bibinfo{volume}{234-236}},
  \bibinfo{pages}{605 } (\bibinfo{year}{1997}).

\bibitem[{\citenamefont{Hordequin et~al.}(2000)\citenamefont{Hordequin,
  Ristoiu, Rannoa, and Pierre}}]{HRR00}
\bibinfo{author}{\bibfnamefont{C.}~\bibnamefont{Hordequin}},
  \bibinfo{author}{\bibfnamefont{D.}~\bibnamefont{Ristoiu}},
  \bibinfo{author}{\bibfnamefont{L.}~\bibnamefont{Rannoa}}, \bibnamefont{and}
  \bibinfo{author}{\bibfnamefont{J.}~\bibnamefont{Pierre}},
  \bibinfo{journal}{Eur. Phys. J. B} \textbf{\bibinfo{volume}{16}},
  \bibinfo{pages}{287} (\bibinfo{year}{2000}).

\bibitem[{\citenamefont{Mavropoulos et~al.}(2004)\citenamefont{Mavropoulos,
  Sato, Zeller, Dederichs, Popescu, and Ebert}}]{MSZ04}
\bibinfo{author}{\bibfnamefont{P.}~\bibnamefont{Mavropoulos}},
  \bibinfo{author}{\bibfnamefont{K.}~\bibnamefont{Sato}},
  \bibinfo{author}{\bibfnamefont{R.}~\bibnamefont{Zeller}},
  \bibinfo{author}{\bibfnamefont{P.~H.} \bibnamefont{Dederichs}},
  \bibinfo{author}{\bibfnamefont{V.}~\bibnamefont{Popescu}}, \bibnamefont{and}
  \bibinfo{author}{\bibfnamefont{H.}~\bibnamefont{Ebert}},
  \bibinfo{journal}{Phys. Rev. B} \textbf{\bibinfo{volume}{69}},
  \bibinfo{pages}{054424} (\bibinfo{year}{2004}).

\bibitem[{\citenamefont{Chioncel et~al.}(2003)\citenamefont{Chioncel,
  Katsnelson, de~Groot, and Lichtenstein1}}]{CKG03}
\bibinfo{author}{\bibfnamefont{L.}~\bibnamefont{Chioncel}},
  \bibinfo{author}{\bibfnamefont{M.~I.} \bibnamefont{Katsnelson}},
  \bibinfo{author}{\bibfnamefont{R.~A.} \bibnamefont{de~Groot}},
  \bibnamefont{and} \bibinfo{author}{\bibfnamefont{A.~I.}
  \bibnamefont{Lichtenstein1}}, \bibinfo{journal}{Phys. Rev. B}
  \textbf{\bibinfo{volume}{68}}, \bibinfo{pages}{144425}
  (\bibinfo{year}{2003}).

\end{thebibliography}
\end{document}